\documentclass[12pt]{wlscirep}
\usepackage{bm,textcomp,enumerate}
\usepackage{bm,bbm,graphicx,epstopdf}
\usepackage{fancyhdr}
\usepackage{subfigure,wrapfig,color}
\usepackage{amssymb,amsfonts,amsmath,psfrag,latexsym,gensymb}
\usepackage{rsphrase,cancel,soul,fancybox}
\usepackage{url}
\hypersetup{colorlinks,citecolor=blue,linkcolor=red,urlcolor=blue}
\AtBeginDocument{\RequirePackage{lmodern, times}}

\def \kbt{$k_{\rm B}T$}

\def \ensmb{$N$-\apt{}-$T$}
\def \sinfo {{\it supplementary information}}
\def \pmf{${\cal W}_{\rm t}$}
\def \epmf{{\cal W}_{\rm t}}
\def \ftet{${\cal F}_{\rm t}$}
\def \eftet{{\cal F}_{\rm t}}
\def \rtet{${\cal R}_{\rm t}$}
\def \ertet{{\cal R}_{\rm t}}
\def \ltet{${l}_{\rm t}$}
\def \eltet{{l}_{\rm t}}
\def \Ltet{${\cal L}_{\rm t}$}
\def \eLtet{{\cal L}_{\rm t}}
\def \sigt{$\sigma$}
\def \esigt{\sigma}
\def \lpt{${\cal L}_{\rm patch}$}
\def \elpt{{\cal L}_{\rm patch}}
\def \apt{${\cal A}_{\rm patch}$}
\def \eapt{{\cal A}_{\rm patch}}
\def \aex{${\cal A}_{\rm ex}$}
\def \eaex{{\cal A}_{\rm ex}}
\def \eaexd{{\cal G}}
\def \kbt{${k}_{\rm B}T$}
\def \ekbt{{k}_{\rm B}T}

\def \mum{\,$\mu$m}

\def \rbead{${\cal R}_{\rm bead}$}
\def \erbead{{\cal R}_{\rm bead}}

\newcommand{\kap}[1]{$\kappa=#1\,k_{\rm B}T$}
\newcommand{\exarea}[1]{${\cal A}_{\rm ex}\sim #1\%$}
\newcommand{\vrbead}[1]{${\cal R}_{\rm bead}= #1$ nm}

\title{Excess area dependent scaling behavior of nano-sized membrane tethers}
\author[1]{N. Ramakrishnan}
\author[2]{Arpita Roychoudhury}
\author[1,3]{David M. Eckmann}
\author[4]{Portnovo S. Ayyaswamy}
\author[5]{Tobias Baumgart}
\author[6]{Thomas Pucadyil}
\author[2]{Shivprasad Patil}
\author[7]{Valerie M. Weaver}
\author[8,9,*]{Ravi Radhakrishnan}

\affil[1]{Department of Bioengineering, University of Pennsylvania, Philadelphia, PA, 19104, USA,}
\affil[2]{Department of Physics, Indian Institute of Science Education and Research, Pune, 411008, India,}
\affil[3]{Department of Anesthesiology and Critical Care, University of Pennsylvania, Philadelphia, PA, 19104, USA,}
\affil[4]{Department of Mechanical engineering and Applied Mechanics, University of Pennsylvania, Philadelphia, PA, 19104, USA,}
\affil[5]{Department of Chemistry, University of Pennsylvania, Philadelphia, PA, 19104, USA,}
\affil[6]{Department of Biology, Indian Institute of Science Education and Research, Pune, 411008, India,}
\affil[7]{Department of Surgery and Anatomy, University of California San Francisco, San Francisco, CA, 94143, USA,}
\affil[8]{Department of Chemical and Biomolecular engineering, University of Pennsylvania, Philadelphia, PA, 19104, USA,}
\affil[9]{Department of Biochemistry and Biophysics, University of Pennsylvania, Philadelphia, PA, 19104, USA}
\affil[*]{\href{rradhak@seas.upenn.edu}{rradhak@seas.upenn.edu}}

\begin{abstract}
Thermal fluctuations in cell membranes manifest as an excess area (\aex{}) which governs a multitude of physical process at the sub-micron scale. We present a theoretical framework, based on an in silico tether pulling method, which may be used to reliably estimate \aex{} in live cells. The tether forces estimated from our simulations compare well with our experimental measurements for tethers extracted from ruptured GUVs and HeLa cells. We demonstrate the significance and validity of our method by showing that all our calculations along with experiments of tether extraction in 15 different cell types collapse onto two unified scaling relationships mapping tether force, tether radius, bending stiffness $\kappa$, and membrane tension $\sigma$. We show that \rbead{}, the size of the wetting region, is an important determinant of the radius of the extracted tether, which is equal to $\xi=\sqrt{\kappa/2\sigma}$ (a characteristic length scale of the membrane) for $\erbead{}<\xi$, and is equal to \rbead{} for $\erbead>\xi$. We also find that the estimated excess area follows a linear scaling behavior that only depends on the true value of \aex{} for the membrane, based on which we propose a self-consistent technique to estimate the range of excess membrane areas in a cell. \\

\end{abstract}

\begin{document}
\maketitle

The mechanical properties of a cell can be used as a surrogate marker to identify cellular phenotypes. Mechanical characterization (or mechanotyping) has been particularly useful in identifying a number of pathophysiologies --- well known examples include the stiffening of malaria infected erythrocytes and hepatocytes, the softening of metastatic cancer cells, and the sickle shape of an erythrocyte laden with hemoglobin S~\cite{Suresh:2007jz,Agus:2013hd,Steward:2013gu}. Several works in biomechanics have aimed to characterize cells based on mechanical measurements using a wide range of techniques such as flow and optical cytometry, manipulation using micropipette aspiration, optical tweezers and laser traps, and microfluidic devices (see~\cite{Suresh:2007jz,Lee:2007hr,VanVliet:2003bo} for comprehensive reviews). These studies have focused on whole cell measurements and hence have investigated the relationship between the mechanotype and pathophysiology at the cellular and tissue scales. In many cases, the changes in mechanical properties are primarily caused by variations in the structure and organization of the cellular cytoskeleton~\cite{Sheetz:2006fg} and the extracellular matrix~\cite{Acerbi:2015by}. Such subcellular scale rearrangements can significantly impact the mechanical properties of the cell membrane at length-scales smaller than cellular dimensions (i.e., tens of nanometers to less than one micron), a range which also corresponds to the scale at which the cell membrane is effective as an organizer and a host of functional signaling complexes.

The  sub-cellular scale relevant to the above discussion corresponds to the dimensions primarily set by the cortical cytoskeletal mesh, which has been estimated to be between $l_c=150-500$ nm~\cite{Ritchie:2003vr,Morone:2006cr}. The mechanical properties of a patch of the cell membrane that spans the region between multiple cytoskeletal pinning points, with typical dimensions $l_c$, can differ from the bulk because the nature of the thermal undulations (and the associated conformational entropy of the membrane) depends directly on $l_c$, and in turn influences the system's free energy. The total area of the membrane (denoted by ${\cal A}$) is in general larger than the projected area of the cytoskeletal mesh (denoted by $\eapt{}$). The characteristics of the membrane deformations and undulations can be described by a dimensionless scalar quantity called the membrane excess area given as $\eaex=100*({\cal A}-\eapt{})/\eapt{}$ and the membrane is taken to be flat when \aex{}=0 and curved/ruffled if $\eaex{}>0$. The presence of excess area (and curvature gradients) can alter the local signaling microenvironment for a number of biophysical processes  whose downstream components include curvature sensing proteins like BAR, Exo70, and ENTH domains~\cite{McMahon:2005km,Zimmerberg:2005jk,Zhao:2013hi}. Notable processes where  modulations in the membrane excess area at the sub-cellular scale can significantly impact common cellular functions including intracellular transport of cargo or viral/bacterial internalization through exo-/endo-/phago-cytosis~\cite{Goh:2013bx,Grant:2009kn}, cell  polarization~\cite{Bryant:2008eb,Orlando:2009cn}, and cell motility~\cite{Luo:2013db}. Hence it is logical to posit that the primary mechanisms linking the cell-microenvironment to cell fate can revolve around the physical factors impacting the membrane at length-scales below $l_c$~\cite{Sheetz:2006fg,Sheetz:2001ed,DizMunoz:2013bi,Paszek:2014it,Miaczynska:2013dj}.

We note that a number of experimental studies have focused on how membranous reservoirs respond to perturbations in the physical environment of the cell. The estimates for excess membrane area determined using conventional morphometric measurements, involving osmotic shock assays and cryo-EM~\cite{SchmidSchonbein:1980ua} do not delineate thermally undulating excess areas, which causes a mis-estimation of the area. Moreover, such methods, by averaging over the entire cell (or even 100s of cells), ignore the heterogeneity on the scale of $l_c$ at a single cell level or the asymmetry in membrane response that could exist in a polarized cell (where the basal and apical surfaces may sustain very different membrane properties). In this article, we propose a theoretical framework/computational model applicable to tether pulling assays (reviewed in ~\cite{Sheetz:2001ed}) to obtain reliable estimates for the membrane excess area. Unique to our modelling approach is a new methodology that allows incorporation of large deformations as well as thermal membrane undulations in the estimate.

\section{Computational model}
We consider a square frame with a lateral size $\elpt{}=510$ nm, which houses the membrane surface. As noted in the introduction ${\cal A}$, \apt{}, and \aex{} are respectively the curvilinear, projected, and excess areas of the membrane. We discretize the membrane surface into a triangulated surface that contains $M$ triangles intersecting at $N$ vertices and forming $L$ links~\cite{Baumgartner:1990fz,Kroll:1992vh} and the statistical weights of the membrane conformations are governed by the discrete form of the Canham-Helfrich Hamiltonian~\cite{Canham:1970wx,Helfrich:1973td}:
\begin{equation}
{\cal H}=\sum \limits_{i=1}^{N} \left\{ \frac{\kappa}{2}\left(c_{1,i}+c_{2,i} \right)^{2}+\esigt{} \right\} {\cal A}_{v}.
\label{eqn:helfrich}
\end{equation}
$\kappa$ and \sigt{} are respectively the bending rigidity and the bare surface tension of the membrane and ${\cal A}_v$ is the curvilinear area per vertex on the surface. $c_{1,i}$ and $c_{2,i}$ are the principal curvatures at a given vertex $i$ computed as in our earlier work~\cite{Ramakrishnan:2010hk}. In our studies we hold \apt{} to be a constant and take $\sigma=0$. However when thermal undulations are taken into account, the effective surface tension in the  membrane will be non-zero due to renormalization effects and a mapping between the renormalized tension and excess area has been quantified in our earlier work~\cite{Tourdot:2014wh}. All our simulations have been performed in a constant \ensmb{} ensemble, where $T$ is the absolute temperature.

The conformational states of the triangulated surface are evolved using the dynamically triangulated Monte Carlo (MC) technique which consists of two independent MC moves: (i) {\it a vertex move} that simulates thermal fluctuations and (ii) {\it a link flip} that captures the fluid nature of biological membranes (see \sinfo{} Sec. S1 for details). A MC step consists of $N$ vertex moves and $L$ link flips that are performed at random and all the moves are accepted using the Metropolis scheme~\cite{Metropolis:1953in}. All the simulations reported here have been performed using a membrane patch with $N=2601$ vertices and the statistics are collected over 1.5 million MC steps.

\subsection{Analytical model for the membrane excess area}  \label{sec:analytic-aex}
The excess area of a planar membrane in the small deformation limit ($| \nabla h | \ll 1$) can be analytically estimated to be~\cite{Helfrich:1984ht,Waheed:2009hf};
\begin{equation}
\eaexd=\dfrac{100}{2{\cal L}_{\rm patch}^2}\sum \limits_{q=q_{\rm min}}^{q=q_{\rm max}}\dfrac{\ekbt{}}{\kappa q^{2}+\esigt},
\label{eqn:aex-smallslope}
\end{equation}
where $q$ denotes the wavenumber of all possible undulation modes in the membrane and  $k_{\rm B}$ the Boltzmann constant. The maximum value of the wavenumber $q_{\rm max}=2\pi a_0^{-1}$ is set by the size of the triangulated vertices $a_0$ and its minimum value $q_{\rm min}=2\pi l_{p}^{-1}$ is set by the length scale $l_p$ such that $l_p \gg a_0$ and $l_p \leq \elpt{}$. We have performed all our analysis using three values of $l_p=150$, $250$, and $510$ nm that represent the variations in the cytoskeletal length-scales. We note that this model only has applicability in the regime of small \aex{} when $| \nabla h | \ll 1$ is satisfied and is expected to fail in regimes where the \aex{} of the cell is not small (see \sinfo{} Sec. S3) .
\subsection{In silico tether pulling assay} \label{sec:tether-pull}
\begin{figure*}
\centering
\includegraphics[width=15cm,clip]{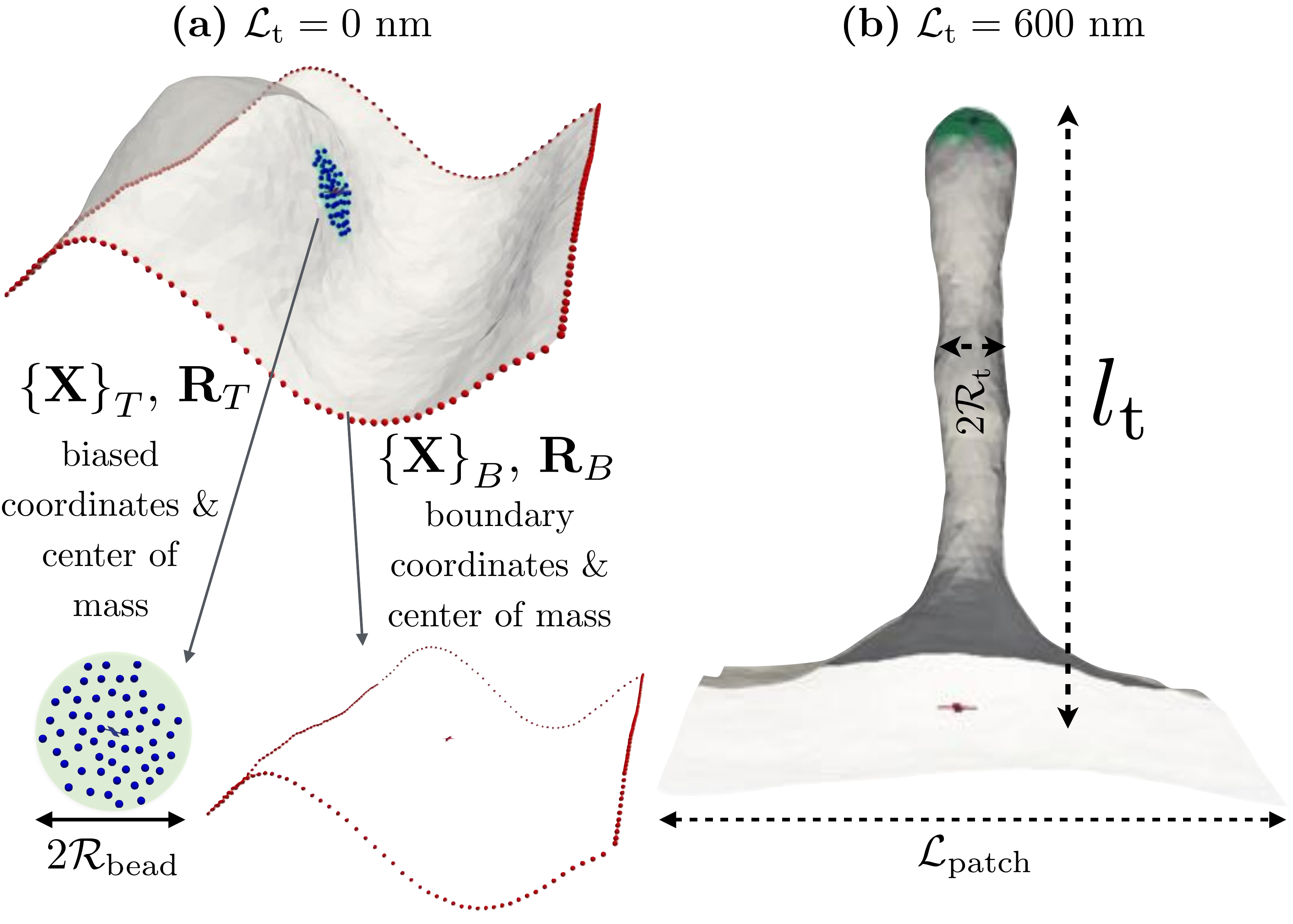}
\caption{\label{fig:tether-illus} (a) Representative equilibrium conformation of a membrane with \kap{20} and \exarea{40}. The set of biased vertices at the tip ($\{{\bf X}_T\}$) and at the base ($\{{\bf X}_B\}$) along with the position of their respective centers of mass ${\bf R}_T$ and ${\bf R}_B$ (shown as crosses) are also shown. $\{{\bf X}_T\}$ is the set of all vertices within a region of size \rbead{}. (b) Conformation of the membrane in panel (a) with a fully developed tether, obtained for $\eLtet{}=600$ nm. The tether force and radius, \ltet{} and \rtet{} and the membrane dimension \lpt{} are also marked.}
\end{figure*}

If \ftet{} be the force required to extract a tether of radius \rtet{} and length \ltet{} from the membrane  patch, as illustrated in Fig.~\ref{fig:tether-illus}, the total energy ${\cal H}_{\rm tot}$, which has a contribution due to membrane deformations (eqn.~\eqref{eqn:helfrich}) and an additional part from the work done to extract the tether (assuming that the tether is a perfect cylinder and ignoring thermal undulations), is given by~\cite{Phillips:2009wv}:
\begin{equation}
{\cal H}_{\rm tot}= \dfrac{\kappa\pi \eltet{}}{\ertet{}}+2\pi\esigt{}\eltet{}\ertet{}-\eftet{}\eltet{}.
\label{eqn:energy-balance}
\end{equation}
Minimization of  the total energy with respect to \ltet{} and \rtet{} yields: (i) $\kappa=\eftet{}\ertet{}/(2\pi)$  and (ii) $\esigt=\eftet{}/(4\pi\ertet{})$. These relationships allow one to determine the elastic properties of the cell membrane through tether pulling experiments; however, the non-trivial geometry of a tether (which in general is not a perfect cylinder) and the underlying membrane patch (which is not a perfect planar entity but rather a ruffled surface subject to undulations, especially under high \aex{}) limits the applicability of eqn.~\ref{eqn:energy-balance}. To overcome these limitations, we have extended the umbrella sampling technique~\cite{Frenkel:2001} to extract tethers of a specified length \Ltet{} from a membrane in the \ensmb{} ensemble. This is analogous to tether extraction in experiments where a constant outward force is applied on a selected region of the  cell membrane through an AFM or an optical tweezer. In our model, we use an additional harmonic biasing potential of the form ${\cal H}_{\rm bias}=k_{\rm bias}(\eltet-\eLtet)^2/2$ in place of the force employed in experiments.  Here $k_{\rm bias}$ is the spring constant of the biasing potential and  \Ltet{} is a reaction coordinate that denotes the prescribed length of the extruded tether. In our calculations we take   $k_{\rm bias}=0.5\,\ekbt{}/{\rm nm}^2$ and this value is chosen such that the undulation modes of the membrane remains unaltered. It should be noted that the addition of the biasing potential does not alters the equilibrium characteristics of the membrane since its contribution will be removed in the WHAM analysis.  

The length of the tether \ltet{} is defined using a macroscopic order parameter, determined from two different sets of vertices $\{{\bf X}_T\}$ and $\{{\bf X}_B\}$, that are shown in Fig.~\ref{fig:tether-illus}(a). ${\bf R}_T$ and ${\bf R}_B$, which are also shown in Fig.~\ref{fig:tether-illus}(a), represent the centers of mass of the chosen vertices that define the two macroscopic variables from which the instantaneous tether length is calculated as $l_t=|{\bf R}_T-{\bf R}_B|$. While $\{{\bf X}_T\}$ is predetermined at the start of the simulation, $\{{\bf X}_B\}$ is computed at runtime and taken to be the set of all vertices at the boundary of the membrane patch (also see \sinfo{} Movie M1).

In a typical tether pulling assay, the bead used to extract the tether is only partially wetted by the membrane surface and in general the wetting area is unknown. Also, due to the non-specific nature of these adhesions the wetting area may vary in different experiments, even for the same cell. In order to investigate the role of the wetting area on the properties of the extracted tether, we choose the biased vertices in the tip to be a circular region of radius \rbead{}. This is illustrated in the lower panel of Fig.~\ref{fig:tether-illus}(a).

\subsection{Potential of mean force}
For a given membrane patch, independent simulations are performed to extract tethers within a given umbrella sampling window. For all simulations reported in this article, we use at least $64$ windows each of width $5$ nm --- the number of windows required to extract fully developed tethers increases with increasing \aex{}. Histograms of the instantaneous tether length in each of the windows are recorded for $1.5$ million Monte Carlo steps and these statistics are converted to a potential of mean force (PMF) using the Weighted Histogram Analysis method~\cite{Roux:1995vi}. The typical runtime for an umbrella-sampling window to sample $1.5$ million MCS is around $36$ hours on a $2.6$ GHz processor.
\subsection{Computing the radius and length of membrane tethers} \label{sec:rad-length-tether}
The radius and length of the membrane tether \rtet{} and \ltet, respectively, can be determined exactly in the simulations, as shown in Fig.~\ref{fig:tether-illus}(b). Let ${\bf [r]}$ be the set of all $N_c$ vertices on the tubular region and ${\bf r}_{CM}=(N_c)^{-1}\sum_i {\bf r}_i$ their center of mass: here ${\bf r}_i$ is the three-dimensional position vector of vertex $i$ in the Cartesian coordinates. The center of mass can be used to construct the gyration tensor as,
${\bf G}=(N_c)^{-1}\sum_{i=1}^{N_c} ({\bf r}_i-{\bf r}_{CM}) \otimes ({\bf r}_i-{\bf r}_{CM})$ whose eigenvalues are  $\lambda_1$, ${\lambda_2}$, and ${\lambda_3}$. Since the tethers formed are axi-symmetric we identify $\lambda_2$ and $\lambda_3$ using the relation $\lambda_2 \approx \lambda_3$. Of the three eigenvalues, $\lambda_1$ represents the length of the tether, with $\eltet{}\approx 2\sqrt{\lambda_1}$, and $\sqrt{\lambda_2}$ and $ \sqrt{\lambda_3}$ represent its two principal radii. We estimate the average tether radius as $\ertet{}=(\sqrt{\lambda_2}+\sqrt{\lambda_3})/2$.
\section{Experimental Methods}
\subsection{Cell culture}
HeLa cells were placed in $35$ mm petridishes at $37$\degree{} C in $5$\% CO$_{2}$ in DMEM (Dulbecco's Modified Eagle's medium, Lonza) containing $10$\% FBS (Fetal Bovine Serum, Gibco) and $0.02$\% Penicillin/Streptomycin for $48$ hours before commencing the experiment. A confluent culture of HeLa cells was treated with $0.25$\% Trypsin-EDTA (Gibco), detrypsinised in DMEM containing $10$\% FBS and seeded at a density of $80,000$ cells/coverslip (Ted Pella Inc., Redding), so that a single monolayer of cells are obtained on the coverslip.

\subsection{Giant Unilamellar Vesicles (GUVs)}
For the preparation of vesicles, $1,2$-dioleolyl-sn-glycero-$3$-phosphocholine (DOPC), $1,2$-dioleolyl-sn-glycero-$3$-phospho-L-serine (DOPS) (Avanti Polar, Alabaster, AL) and $1,2$-dioleolyl-sn-glycero-$3$-phosphoethanolamine-N-(lissamine rhodamine B sulfonyl)(RhPE) (Invitrogen) stock solutions in chloroform, at room temperature were used. The lipid mix was aliquoted in a glass vial to a total lipid concentration of 1 mM at a ratio of DOPC:DOPS:RhPE ($84$:$15$:$1$ mol\%).

Gel-assisted formation of GUVs were carried out using polyvinyl alcohol (PVA) as described earlier~\cite{Weinberger:2013gq}, with a few modifications as per the requirements of the experiments. In this method of GUV formation, a drop of $5$\% w/v degassed PVA (MW $145,000$, Sigma) in deionized water is added to a clean glass coverslip placed on a hot plate set at $75$\degree{} C. The water gets evaporated in about $10$ minutes leaving a dry thin film of PVA on the coverslip. To this, around $3$ $\mu$L of the $1$ mM lipid stock solution in chloroform was added to dry PVA while on the hot plate to let the chloroform evaporate. The thin film was peeled off and immersed in eppendorfs containing $20$ mM HEPES, $150$ mM NaCl, pH $7.4$ with $100$ mM sucrose. This immersed film was left undisturbed for around one hour followed by gentle tapping to release the GUVs from the PVA film to the buffer solution. The buffer containing large free floating GUVs ($10$-$15$ $\mu{\rm m}$) was pipetted out and used for tether pulling experiments. 

\subsection{AFM Experiments}

AFM-based force spectroscopic experiments were performed using Nanowizard II atomic force microscope (JPK Instruments).  The AFM liquid cell was assembled with freshly cleaved mica discs prior to adding the GUV solution. The liquid cell was then mounted on the AFM stage and left undisturbed for $20$ minutes to allow the vesicles to settle on the mica surface. Using a fluorescence microscope attached with the AFM set up, we could confirm that the GUVs settled on the surface and the floating ones were washed away by exchanging buffer solution with HBS. Subsequently, the GUVs got ruptured on the mica surface and they were imaged using AFM. The images obtained using AFM revealed the location and height of the ruptured GUV patches which matched with that of the height of a single bilayer membrane ($5$-$6$ nm). Force spectroscopy was then performed on these particular patches to pull membrane tethers. Silicon nitride cantilevers (MikroMasch CSC$38$/AlBS) were used for pulling the tethers. Cantilevers were calibrated before each experiment and its spring constant was determined using equipartition theorem~\cite{Hutter:1993ed}. The measured spring constant of the cantilevers used for most experiments was found to be range of $20$-$80$ mN/m. Constant speed mode was used for approaching the tip to the sample surface followed by retraction at the same speed. The approach-retract cycle was repeated at various points on the membrane patch using force mapping tool built in Nanowizard II software and force-displacement curves were recorded. Force curves showing step  profiles were selected and analyzed using JPK data processing software by fitting the curves with the in-built functions to measure the force minimum corresponding to the tether force and step heights in retraction force curves. 
\section{Results}
\subsection{Extraction of membrane tether proceeds through three distinct regimes}  \label{sec:tetherregimes}
\begin{figure}
	\centering
	\includegraphics[width=15cm,clip]{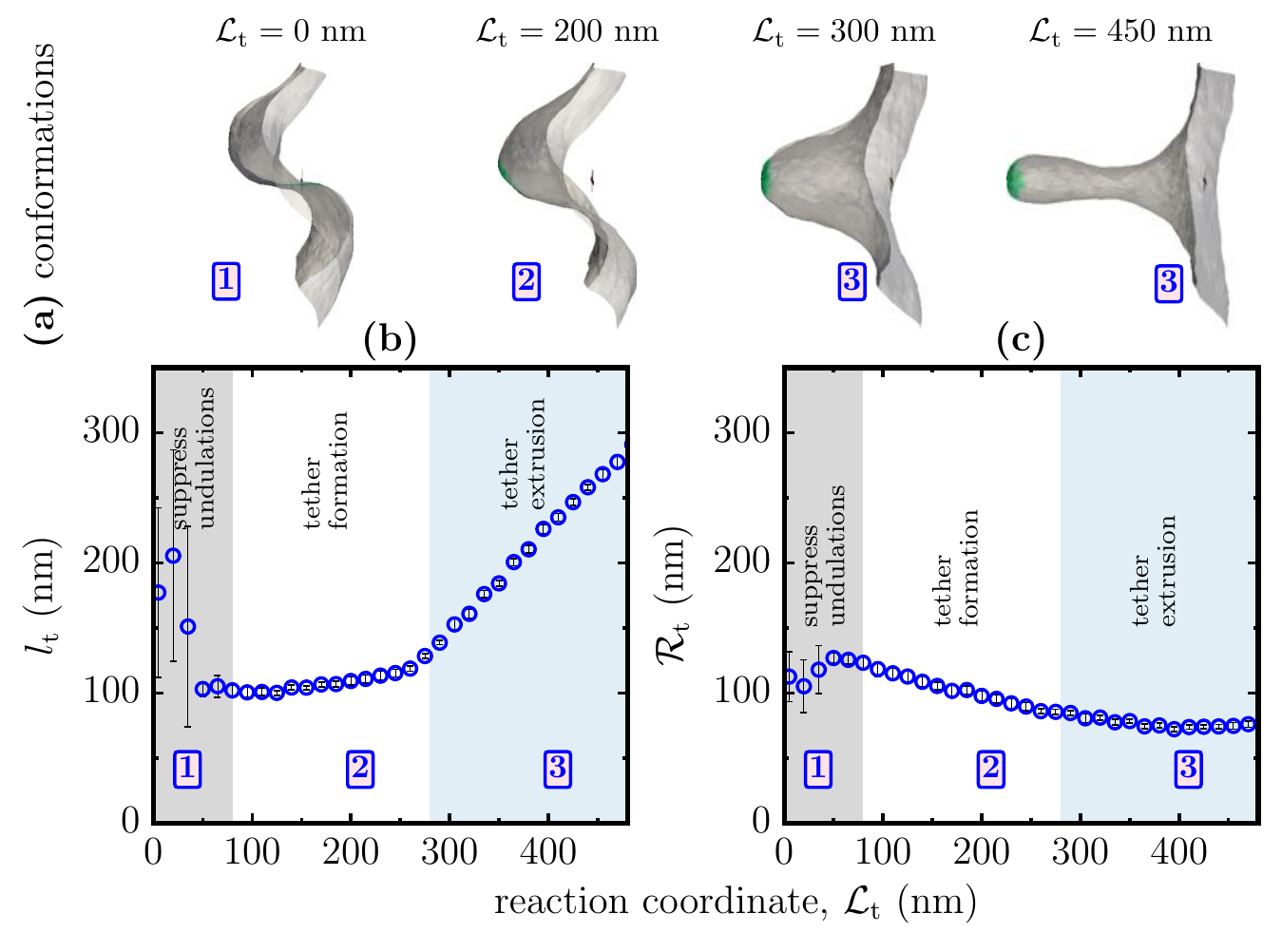}
	\caption{\label{fig:compare-beta} (a) Representative conformations of a membrane with \kap{20} and \exarea{40} as a function of \Ltet{}. Panels (b) and (c) show the computed values of the tether length \ltet{}, and radius \rtet{}, respectively, as a function of \Ltet{}. These quantities are computed as described in Sec.~\ref{sec:rad-length-tether}. The shaded regions mark the three regimes for tether extraction namely, regime \textbf{1}: suppression of undulations, regime \textbf{2}: formation of tethers, and regime \textbf{3}: extrusion of tethers at a constant radius. The boxed numbers in the top panel denote the regimes to which the configurations correspond to.}
\end{figure}

We first demonstrate the characteristics of a tether extracted from a model membrane with $\kappa=20$ \kbt{} and \exarea{40}, using a bead size of \vrbead{50} in the \ensmb{} ensemble. The tether is extracted using the umbrella sampling technique described in the methods section, for reaction coordinate (imposed tether length) values in the range $0<\eLtet<500$ nm, with a window size of $5$ nm. The top panel in Fig.~\ref{fig:compare-beta} shows representative snapshots of the membrane stabilized at four different values of \Ltet{ = $0$, $200$, $300$, and $450$} nm. At small values of \Ltet{}, the membrane conformations show large undulations whose magnitudes are set by the value of \aex{}. However, at large values of \Ltet{}, the membrane undulations are absorbed into the large out of plane protrusions that resemble a tether extracted from a planar membrane. It is noted that the shape of a fully developed tether (i.e., when the undulations in the planar region becomes very small) is consistent with that predicted for nearly planar membranes, using analytical methods~\cite{Derenyi:2002kx}.

The instantaneous length and radius of the tether region, denoted by \ltet{} and \rtet{}, as a function of the reaction coordinate \Ltet{}, are shown in the middle and lower panels of Fig.~\ref{fig:compare-beta}, respectively. Both \ltet{} and \rtet{} show non-monotonic behaviors with respect to \Ltet{}, which are solely attributable to the non-zero excess area of the membrane. For membrane with thermal undulations, and hence non-zero excess areas, we identify three characteristic regimes for tether growth which are marked as shaded regions in the figure. These regions are characterized as follows:
\begin{itemize}
\item \textbf{Regime 1} ($\eltet{} \approx \ertet{}$): for \Ltet{$<75$ nm}, where the tether radius and length are similar, the applied biasing potential only serves to suppress the short wavelength undulations in the membrane. This is reflected in the fact that the membrane conformations in this regime are not distinguishable from their equilibrium counterparts. \item \textbf{Regime 2} ($\eltet{} \approx $ constant and  $\ertet{}\propto \eLtet{}^{-1}$): for $75<\eLtet<300$ nm a pronounced protrusion is seen in the vicinity of the region where the biasing potential is applied. The radius of this protrusion decreases with increasing \Ltet{}, while its length remains unchanged. 
\item \textbf{Regime 3} ($\ertet{} \approx$ constant and $\eltet{} \propto \eLtet{}$): for \Ltet{$>300$ nm} in Fig.~\ref{fig:compare-beta}, the tether radius remains constant while its length increases linearly with \Ltet{}, marking a region of tether growth. The linear increase in \ltet{} fails to hold  when all excess area in the membrane is drawn into the tether region.
\end{itemize}

The extent of the three regimes, depend on the values of $\kappa$ and \aex{}. This is shown in the \sinfo{}, where we have displayed the effects of \aex{} and $\kappa$ on the radius of the extracted tether.

\begin{figure}
\centering
\includegraphics[width=15cm,clip]{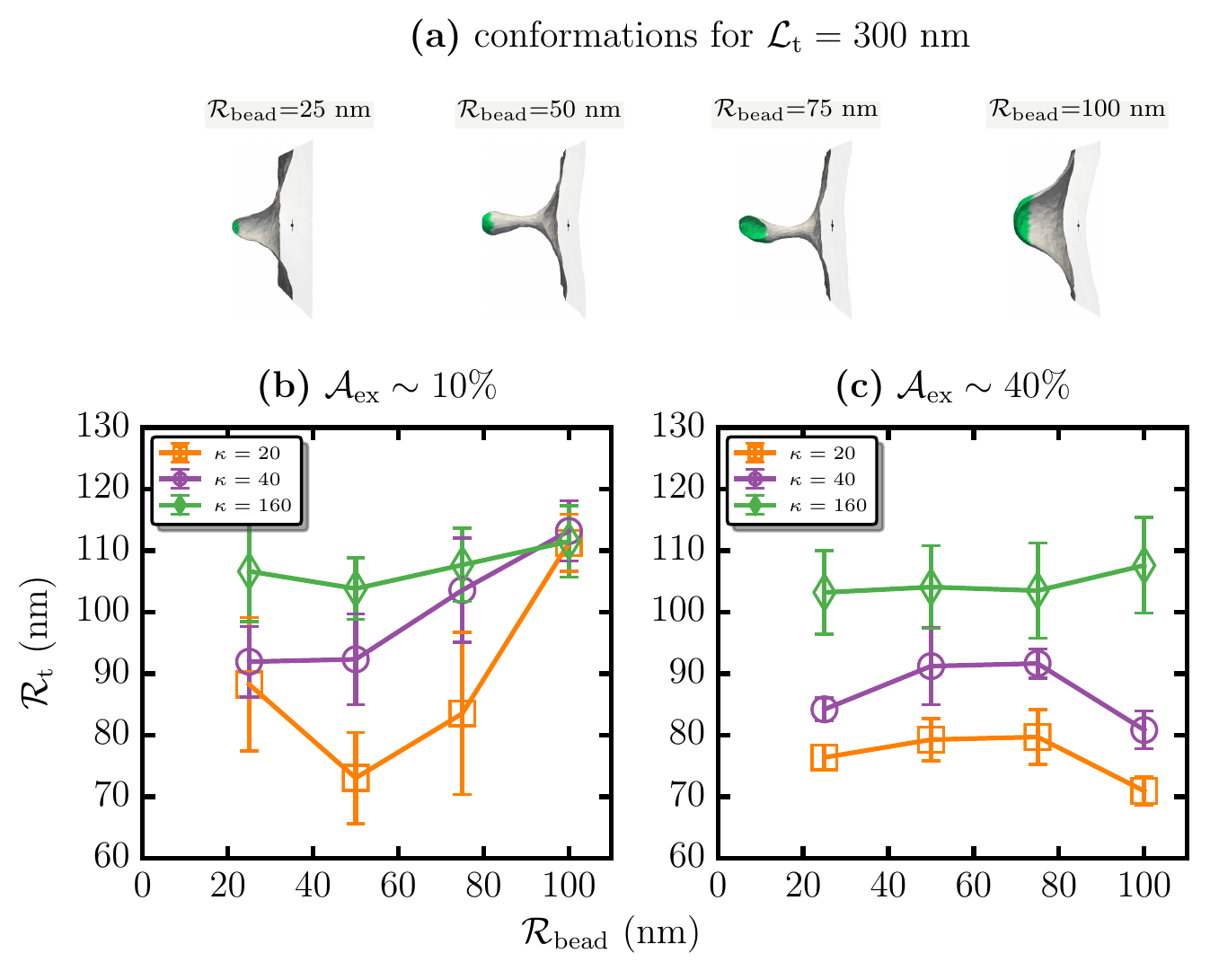}
\caption{\label{fig:beadradius} Dependence of the tether radius on the size of the biasing region. (a) Representative conformations of tethers extracted using beads with $\erbead=25$, $50$, $75$, and $100$ nm, from a membrane with \kap{20} and \exarea{10}. Panels (b) and (c) show the computed values of \rtet{}, as a function of \rbead{}, for $\kappa=20,\,40,$ and $160$ \kbt{} for $\eaex{}=10$ and $40\%$, respectively.}
\end{figure}

The characteristic length scale for a membrane, given by $\xi=\sqrt{\kappa/2\sigma}$~\cite{Lipowsky:1991cu,Seifert:1997wq}, sets the limit below which curvature contributions are dominant. In our model, $\xi$ is an increasing function of $\kappa$ and \aex{} --- the latter may be deduced from the inverse relationship between $\sigma$ and \aex{} in eqn.~\eqref{eqn:aex-smallslope}. In a tether pulling experiment performed in the \ensmb{} ensemble, the radius of the extracted tether depends either on $\xi$ or on the size of the biased region \rbead{} used for tether extraction. This is shown in Fig.~\ref{fig:beadradius} where we display the values of \rtet{} as a function of \rbead{}, for $\kappa=20,\,40,$ and $160$ \kbt{} and $\eaex{}=10$ and $40\%$. The conformations  shown in panel (a) for a membrane with \kap{20} and \exarea{10}, for $\eLtet{=300}$ nm, clearly illustrates the interplay between the characteristic length $\xi$ and the imposed length \rbead{}. While we observe fully grown and geometrically identical tethers for $\erbead \leq 75$ nm, we find the tether extracted with $\erbead = 100$ nm to be significantly different. This feature is also quantified in Fig.~\ref{fig:beadradius}(b) where we find the nearly constant tether radius ($\ertet{} \sim 80$ nm) for $\erbead \leq 75$ nm to show a marked increase to $\ertet{} \sim 110$ nm when $\erbead = 100$ nm. 

 In panels (b) and (c) of Fig.~\ref{fig:beadradius} two key features are worth noting: (i) as expected, the value of \rtet{} is an increasing function of $\kappa$ for all values of \rbead{}, and (ii) the dependence of \rtet{} on \rbead{} is minimal for large values of $\kappa$ and also when \aex{} is large.

\subsection{PMF and tether force} 
\begin{figure}
\centering
\includegraphics[width=15cm,clip]{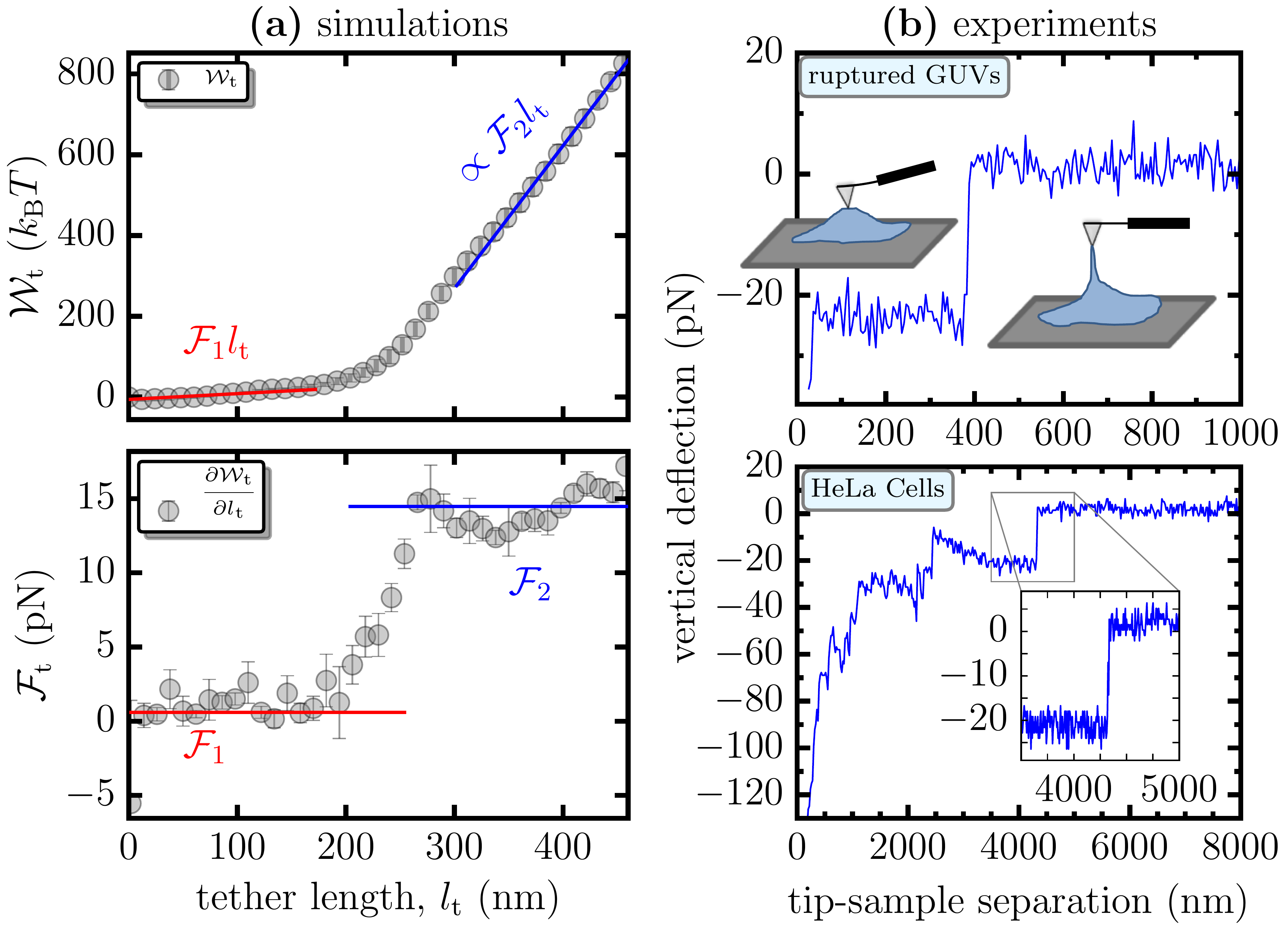}
\caption{\label{fig:rg-pmf-force} (a) The potential of mean force \pmf{} and the tether force \ftet{}, as a function of the tether length \ltet{}, for a membrane with \kap{20} and \exarea{40}. In the top panel, \pmf{} shows a linear scaling in regimes 1 and 3, which are represented by the functions ${\cal F}_1\eltet{}$ and ${\cal F}_2\eltet{}$, respectively. The lower panel compares values of \ftet{} estimated from direct numerical differentiation of \pmf{} (symbols) to that obtained from the scaling relations (lines). (b) Force displacement curves for experimental tether pulling assay using ruptured GUVs (top panel) and HeLa cells (lower panel) -- the inset shows a transition between regions of constant force. The illustration in the top panel shows the state of the membrane tether at various stages of the experiment. The vertical deflection of the AFM tip is measure of the tether force \ftet{} and its separation from the sample is a measure of the tether length \ltet{}.}
\end{figure}

The PMF (\pmf{}) to extract a tether of length \ltet{} from a membrane patch of fixed \aex{} is computed from the umbrella sampling data using the WHAM technique (see methods section). \pmf{} for a membrane with \kap{20} and \exarea{40} is shown in the top panel of Fig.~\ref{fig:rg-pmf-force}(a). The three characteristic regimes seen for \rtet{} (see Sec.~\ref{sec:tetherregimes}) are also reflected in the form of \pmf{}. Here, we again observe three scaling regimes : (i) an initial linear regime given by ${\cal F}_1\eltet{}$, (ii) a second non-linear regime, $\propto\eltet{}^2$, and (iii) a final linear regime, $\propto {\cal F}_2 \eltet{}$. Both the linear regimes are shown as solid lines in panel (a) of Fig.~\ref{fig:rg-pmf-force} and the latter is attributable to tether extrusion at a constant radius, for which the elastic energy is expected to scale as ${\cal H}_{\rm tot} \propto \eltet{}$ (eqn.~\eqref{eqn:energy-balance}). On the other hand, the source of the non-linear scaling is attributed to \rtet{} being a decreasing function of \ltet{}. We note that the scaling behavior is universal and is observed for all systems investigated.

The force required to extract the tether may be computed as $\eftet{}=|\nabla_{\eltet{}}\epmf{}|$, where $\nabla_{\eltet{}}$  denotes a gradient with respect to \ltet{}. \ftet{} can be estimated either from  direct numerical differentiation of \pmf{} or from the scaling relations --- for the latter, $\eftet{}={\cal F}_1$ in regime 1 and $\eftet{}={\cal F}_2$ in regime 3. The tether forces computed using the two methods for \pmf{} in Fig.~\ref{fig:rg-pmf-force}(a) are shown in the lower panel --- symbols and lines correspond to \ftet{} obtained using numerical differentiation and using the scaling relations, respectively. We find the estimates from both the methods to be in excellent agreement. Since direct numerical differentiation is subject to a large noise to signal ratio, we  primarily rely on the scaling relation based method to estimate \ftet{}. As in experiments, we report the  value of the  force in the second regime as the tether force, i.e., $\eftet{} \sim {\cal F}_{2}$.

The tether force shown in Fig.~\ref{fig:rg-pmf-force}(a) has the same qualitative and quantitative behavior as that normally observed in experiments. The top and bottom panels in Fig.~\ref{fig:rg-pmf-force}(b) show forces required to extrude a tether from ruptured GUVs on mica and from the HeLa cells, respectively. The pulling speeds in both the experimental assays are taken to be 1 $\mu$m/s, which satisfies the assumption of quasi-equilibrium tether extraction employed in our simulations. Measurements at speeds less than that reported here are not possible due to the noise arising from cantilever thermal drift.  Though there are no known techniques to calculate the precise value of \aex{} for both systems, it is reasonable to assume that it is finite.
While the force-displacement curves for both the systems depend on the properties of their respective bilayer membrane, in the case of HeLa cells there may be additional contributions due to the underlying cytoskeletal mesh. Though we would expect ruptured GUVs on a mica surface to be free of any pinning contacts, there could be a finite number of pinning sites due to the chemical heterogeneity on the surface in spite of the surface being atomically smooth. The salt concentration in the buffer may screen the interactions between the membrane and the mica surface leading to a sparse contact between the two and the effect of these non-specific contacts on the force-displacement curves are minimal. The forces measured in experiments match very well with the numerically computed values of \ftet{}. The measured tether force is about 20 pN for tethers pulled from both the ruptured GUVs and the HeLa cells. For the case of ruptured GUVs, the tether length at which we observe a transition to the tether extrusion regime is consistent with that seen in our simulations, while that for the cells is considerably higher extending into few microns. We attribute this deviation to the lack of a suitable reference frame for cellular measurements.

\begin{figure}
\centering	\includegraphics[width=15cm,clip]{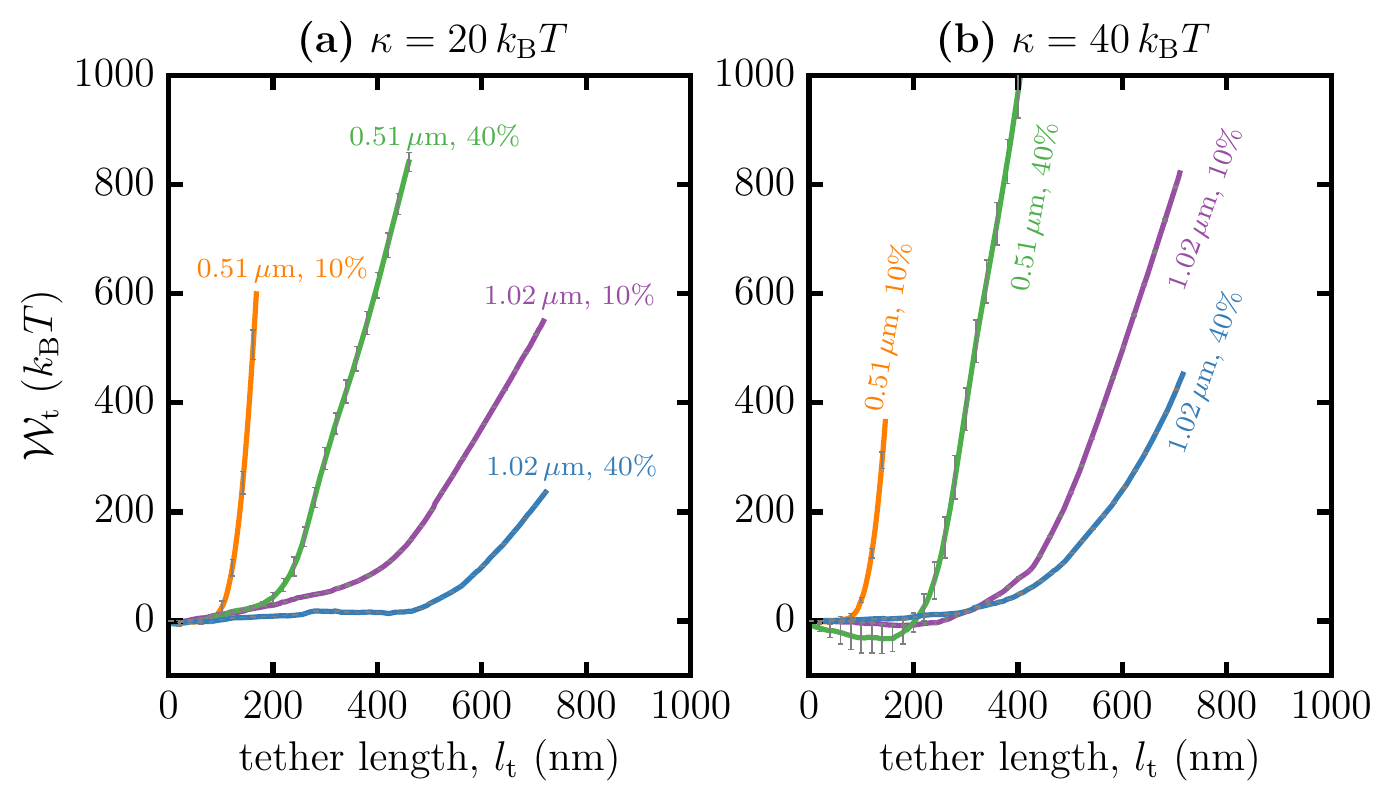}
\caption{\label{fig:pmf-size} The potential of mean force \pmf{} as a function of the tether length \ltet{}, extracted with $\erbead{}=50$ nm,  from membranes with $\elpt{}=0.51$\mum{} and $1.02$\mum{}, and excess areas $\eaex=10\%$ and $40\%$. Data for \kap{20} are shown in panel (a) and that for \kap{40} is shown in panel (b).}
\end{figure}
As noted in the introduction, the size of the cytoskeletal mesh ($l_{c}$) bounding the cell membrane significantly influences the characteristics of the extracted tether. The current theoretical model only considers tethers from a homogeneous membrane with constant $\kappa$ and \aex{}. However, to zeroth order, the role of the cytoskeleton in suppressing long wavelength undulations beyond $l_c$ can be taken into account in our model by examining the dependence on the membrane patch size \lpt{}. In Fig.~\ref{fig:pmf-size}, we investigate this effect by extracting tethers from two planar patches with $\elpt{}=510 $ nm and $\elpt{}=1.02 $ \mum{}, which are representative of cell membranes scaffolded by dense and sparse cytoskeletal meshes, respectively. Panels (a) and (b) show data for membranes with $\kappa=20$ and $40$ \kbt{}, respectively, for excess areas $\eaex{}=10$ and $40\%$. It is evident from these figures that the PMF, and hence \ftet{} and \rtet{}, in addition to the elastic parameters $\kappa$ and \aex{}, are also functions of \lpt{}. This points to the fact the cell may have a heterogeneous mechanical microenvironment depending on the cytoskeletal mesh size and may provide varied response to biochemical processes, such as nanocarrier or viral binding, depending  of the characteristic value of $l_{c}$ at the site of the process~\cite{Ramakrishnan:2016fl}. Hence, characterizing the mechanical properties of the cell membrane at the scale of $l_{c}$ would be extremely important. In the following, we will only focus on membrane patches with $\elpt{}=510$ nm to establish how the excess area of the membrane can be inferred from tether pulling experiments.

\subsection{Tether radii and forces measured \textit{in silico} compare well with range of values measured in \textit{in vivo} experiments} \label{sec:fourtype}
\begin{figure}
	\centering
	\includegraphics[width=15cm,clip]{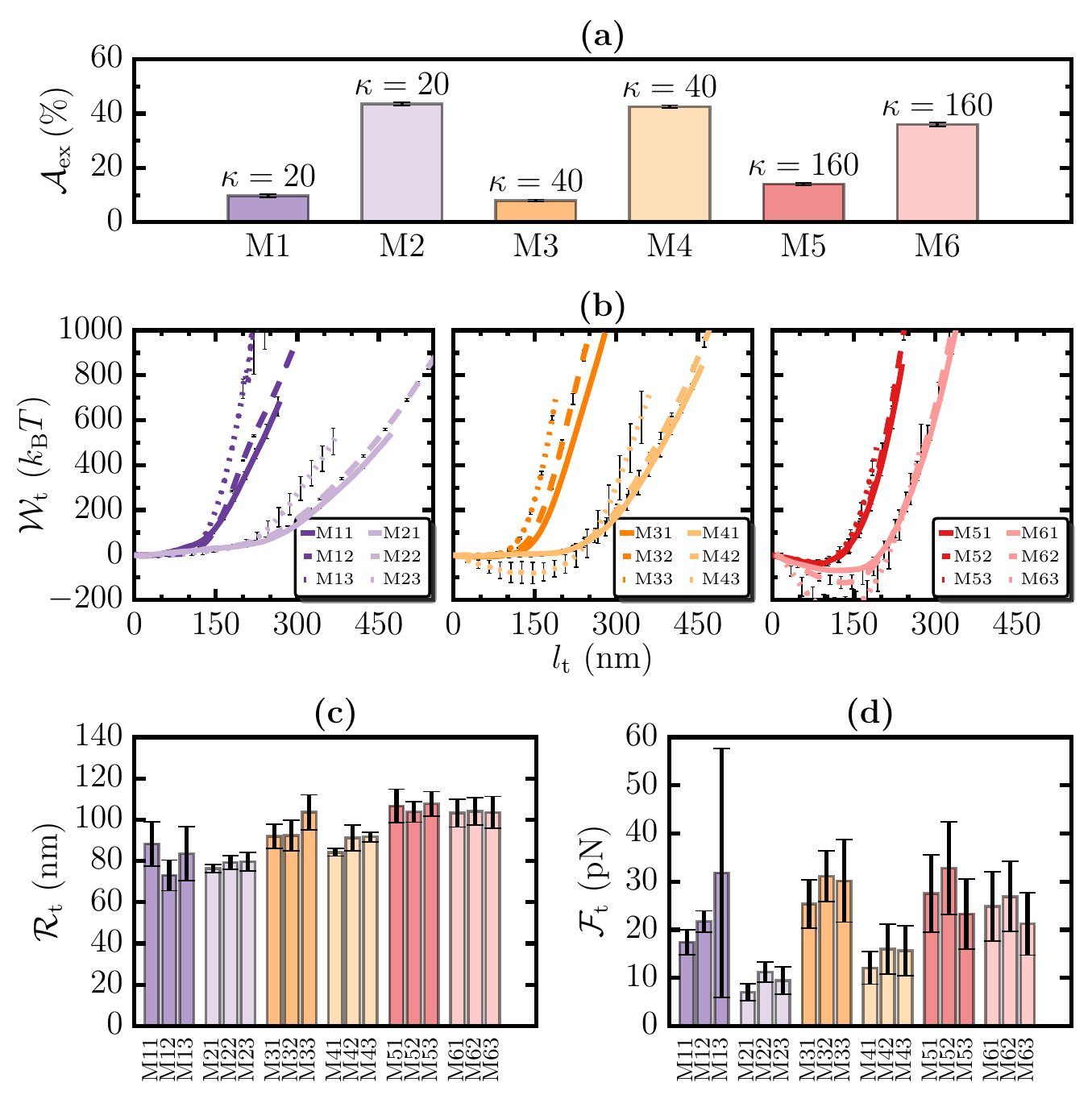}
	\caption{\label{fig:pmf-sixcells}(a) Six model membrane systems, denoted  M1--M6, with specified values of \aex{} and $\kappa$. For any system M$_i$ ($i=1\cdots 6$), M$_{i1}$, M$_{i2}$, and M$_{i3}$ correspond to tethers extracted with $\erbead=25$, $50$, and $75$ nm, respectively. The values of \pmf{}, \ftet{}, and \rtet{} for all the systems are shown in panels (b), (c), and (d), respectively.}
\end{figure}

Pontes et. al.~\cite{Pontes:2013gl} have recently reported results for \textit{in vivo} tether pulling assays studies of 15 different cell types in the central nervous system (CNS) --- the data is also shown in the \sinfo{}. Based on this study, we classify cells in the CNS into four distinct categories: (i)  small $\kappa$ ($20-60$\kbt{}) \& small $\sigma$, (ii)  small $\kappa$ \& large $\sigma$, (iii) large $\kappa$ ($\sim 160$ \kbt{}) \& small $\sigma$, and (iv) large $\kappa$ \& large $\sigma$. In order to establish the quantitative accuracy of our model, we compute the values of \rtet{} and \ftet{} for six model systems which are representative of the cells in the CNS. They are denoted by M1 (\kap{20}, \exarea{10}), M2 (\kap{20}, \exarea{44}), M3 (\kap{40}, \exarea{9}), M4 (\kap{40}, \exarea{43}), M5 (\kap{160}, \exarea{13}), and M6 (\kap{160}, \exarea{38}). These model systems are also depicted in Fig.~\ref{fig:pmf-sixcells}(a).

We extract tethers from all the six model system (M$_i$, with $i=1\cdots 6$), using bead sizes $\erbead=25,\,50$, and $75$ nm --- the corresponding data are denoted by M$_{ij}$, where $j=1$, $2$, and $3$, respectively. The PMFs for these systems are displayed in Fig.~\ref{fig:pmf-sixcells}(b) and the presence of the three characteristic regimes for \pmf{}, discussed earlier, are evident. Despite a similarity in the scaling behavior, the values of \pmf{} are highly sensitive to changes in both \rbead{} and the elastic parameters $\kappa$ and \aex{}, predominantly so for the latter. The average values of \rtet{} and \ftet{} for the model systems are displayed in Figs.~\ref{fig:pmf-sixcells}(c) and (d) respectively. \rtet{} is found to be independent of \rbead{} and, as expected, we find:  (i) for a given $\kappa$, \rtet{} is a decreasing function of  \aex{} (e.g. M1$>$M2), and (ii) for a fixed \aex{}, \rtet{} is an increasing function of $\kappa$   (e.g. M5$>$M3$>$M1). The tether force also shows a similar behavior, with \ftet{} being larger for systems with smaller \aex{} and larger $\kappa$. The range of values for the tether force ($10<\eftet{}<50$ pN) and radius ($60<\ertet{}<110$ nm) measured in our simulations compare very well with the experiments of Pontes et. al.~\cite{Pontes:2013gl}, where they report values in the range $15<\eftet{}<70$ pN  and $43<\ertet{}<158$ nm. This establishes the validity of our present model as a tool for interpreting tether pulling assays that aim to probe tethers in the nanoscopic scale.

\begin{figure}
\centering
\includegraphics[width=15cm,clip]{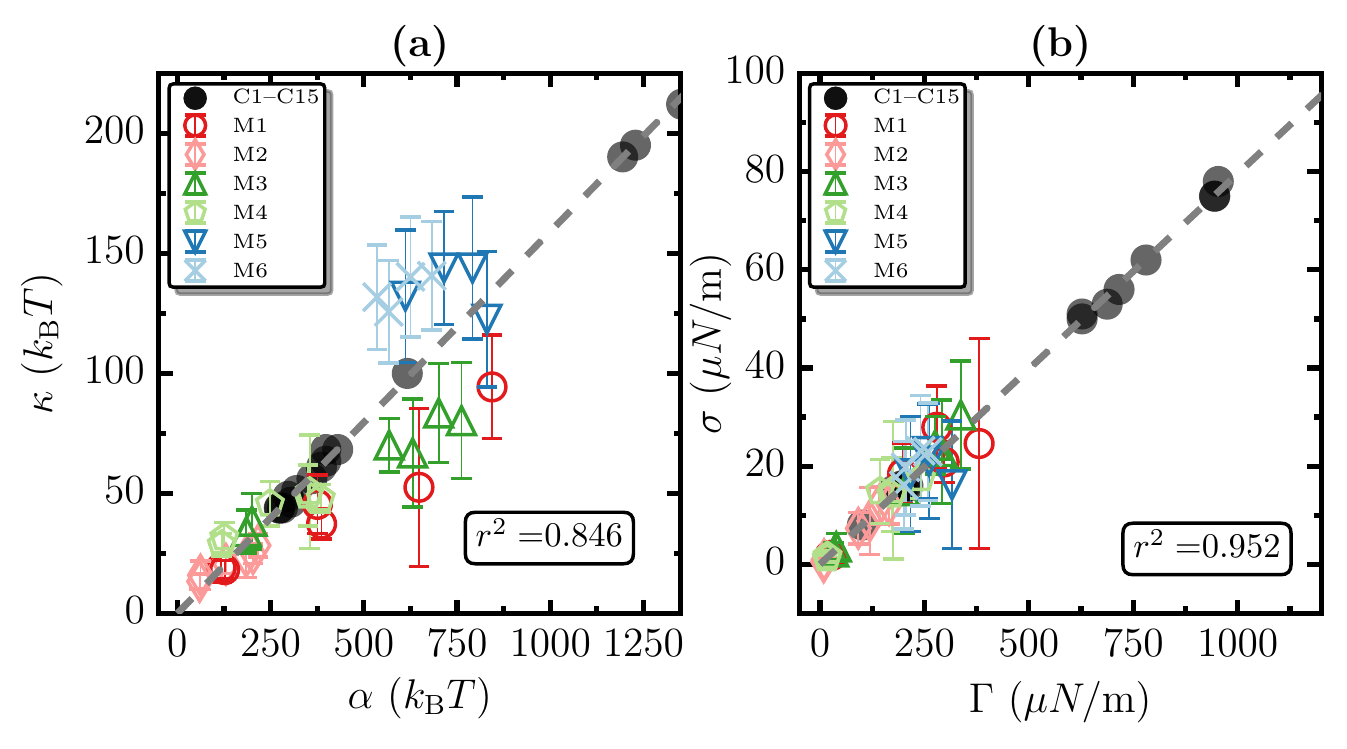}
\caption{\label{fig:kappa-sigma-scaling} Validity of the scaling relations for $\kappa$ and $\sigma$ for data from simulations (M1--M6, shown as open symbols) and experiments (C1--C15, shown as filled symbols). Panel (a) shows the relation $\kappa/\alpha = 1/2\pi$ and panel (b) shows the scaling relation $\sigma/\Gamma=1/4\pi$, and the corresponding correlation coefficients for systems $M1-M6$ are found to be $r^2=0.846$ and $r^2=0.952$, respectively. The dotted lines in panels (a) and (b) correspond to $1/2\pi$ and $1/4\pi$ respectively.}
\end{figure}

Our results in Figs.~\ref{fig:kappa-sigma-scaling}(a) and (b), depict the adherence to the constitutive relations derived by minimizing eqn.~\eqref{eqn:energy-balance}. Briefly, the effective bending rigidity and the surface tension are expected as follow the relations $\kappa/\alpha =(2\pi)^{-1}$ and $\sigma/\Gamma = (4\pi)^{-1}$, respectively. Here the scaling parameters are $\alpha=\eftet{}\ertet{}/\ekbt{}$ and $\Gamma=\eftet{}/\ertet{}$. As can be seen from the figures, data from both our simulations (marked M1--M6 and shown as open symbols) and from the experiments of Pontes et. al.~\cite{Pontes:2013gl} (marked C1--C15 and shown as filled symbols) show a good collapse, with correlation coefficients of $r^2=0.846$ for $\kappa$ and $r^2=0.952$ for $\sigma$, which further establishes the agreement of our calculations and the referred experiments with known scaling relationships. The dotted lines  in Figs.~\ref{fig:kappa-sigma-scaling}(a) and (b) correspond to $(2\pi)^{-1}$ and $(4\pi)^{-1}$, respectively. 

\subsection{Data from tether pulling experiments may be classified according to \aex{}}
\begin{figure}
\centering
\includegraphics[width=15cm,clip]{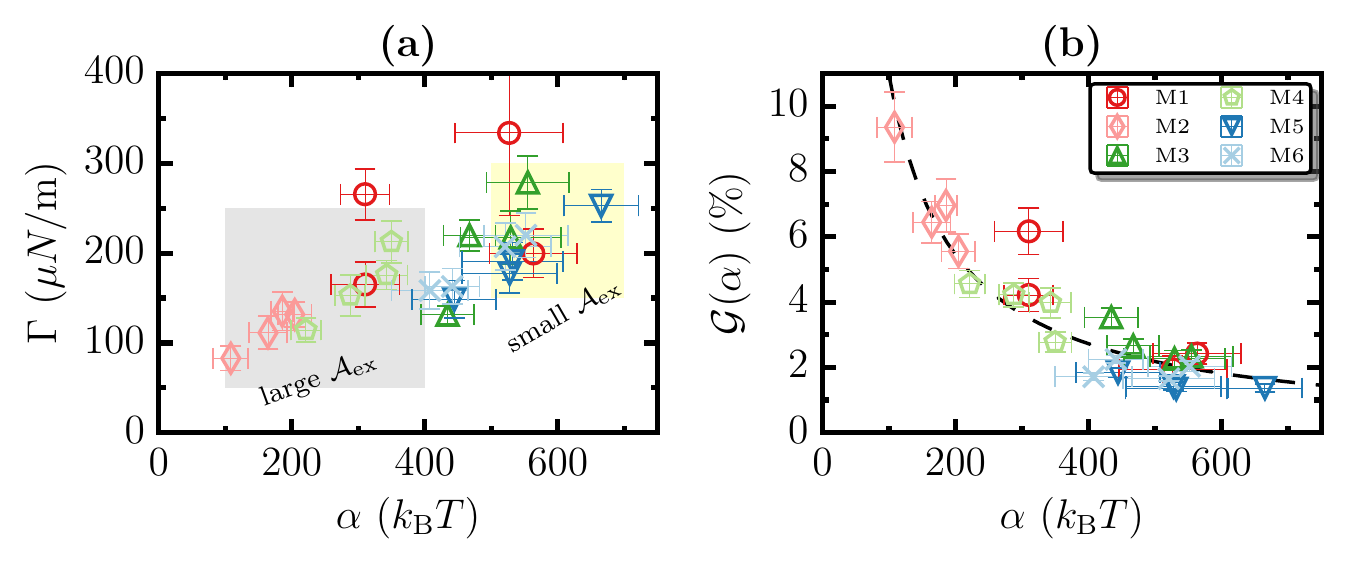}
\caption{\label{fig:G-scaling}  (a) A plot of $\alpha$ vs $\Gamma$ for M1--M6, for different values of \rbead{}, show data clustering in an excess area dependent fashion. (b) ${\cal G}(\alpha)$, the analytical estimates for the membrane excess area for M1--M6, computed using eqn.~\eqref{eqn:aex-smallslope}. The dotted line denotes a scaling of the form $G/\alpha$, with $G\sim 1107$.}
\end{figure}

Using a suitable choice of scaling parameters, data from various tether pulling assays may be classified according to the excess area in the membrane. We demonstrate this feature in Fig.~\ref{fig:G-scaling}(a) where we show a plot of $\alpha$ vs $\Gamma$ for the six model systems we have chosen. Each system is represented by a set of four data points which correspond to tethers extracted with $\erbead=25$, $50$, $75$, and $100$ nm. The entire set of data clusters into groups, that are primarily dependent on the value of \aex{} in the model membrane. It may be seen that systems M1, M3, and M5 (with \exarea{10}) are clustered in the top right while M2, M4, and M6 (with \exarea{40}) are clustered in the bottom left, and these two clusters are marked as shaded regions. Such a clustering analysis provides a useful route to experimentally classify cells. However, it does not yield any information about the value of \aex{}.

Based on eqn.~\eqref{eqn:aex-smallslope}, we recognize that ${\cal G}(\alpha)$ shows a scaling of the form $G/\alpha$ (dotted line in Fig.~\ref{fig:G-scaling}(b)). The data from our calculations are consistent with this scaling as depicted in Fig.~\ref{fig:G-scaling}(b). Given the potential for clustering of our data in Fig.~\ref{fig:G-scaling}(a) on the basis of \aex{}, and the scaling shown in ${\cal G}(\alpha)$ in Fig.~\ref{fig:G-scaling}(b), we define a dimensionless variable $\eta=\eaex{}/{\cal G}$.

\begin{figure}
	\centering
	\includegraphics[width=15cm,clip]{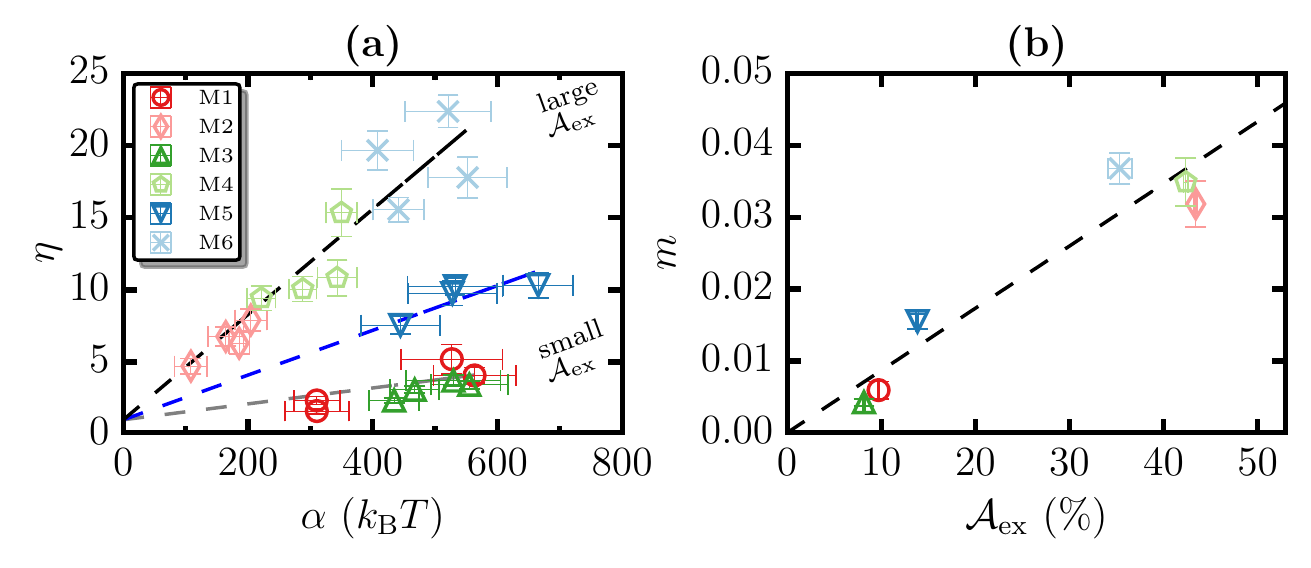}
	\caption{\label{fig:eta-scaling} (a) Scaling plot of $\eta$ vs $\alpha$ for systems M1--M6 for four different values of \rbead{}. The dotted lines, show representative scaling relations of the form $\eta_i = m_i \alpha +1$, for small, intermediate, and large \aex{} regimes. (b) A plot of the slope $m_i$ as a function of \aex{} and the dotted lines denote the best linear fit to the data. Fitting $f(\eaex{}_{,i})=K\eaex{}_{,i}$ we find the value of $K=0.00085/(\ekbt{})$.}
\end{figure}

A plot of $\eta$ as a function of $\alpha$ for systems M1--M6, for four different values of \rbead{}, are shown in Fig.~\ref{fig:eta-scaling}(a). Intriguingly, the data collapse into a linear scaling behavior when $\eta$ is plotted against $\alpha$ (see Fig.~\ref{fig:G-scaling}(a)) where the slope of the scaling line depends only on \aex{}. The scaling is represented as:

\begin{equation} 
\eta_i = m_i \alpha + 1, 
\label{eqn:etascaling}
\end{equation} 
with $i=1\cdots6$. The intercept is taken to be $1$ since $m_i \rightarrow 0$ as $\eta_i \rightarrow 1$, i.e., when ${\cal G} \rightarrow \eaex{}$. We estimate the values of $m_i$ for each system by fitting the corresponding data to a linear function. The three representative dotted lines in Fig.~\ref{fig:G-scaling}(a), corresponding to the small, intermediate, and large excess area regimes, show the clustering of data that only depends on the value of \aex{} in the membrane. The values of $m_i$ computed for each set of data in M1--M6 (Fig.~\ref{fig:eta-scaling}(a)) are shown as a function of \aex{} in Fig.~\ref{fig:eta-scaling}(b). In general, the dependence of $m_i$ on \aex{} may be expressed as:
\begin{equation} 
m_i=f(\eaex{}_{,i}),
 \label{eqn:mifunc}
\end{equation} 
 where $f$ is an unknown function. As a first approximation, we find  $m_i$ to be a linear function of \aex{}  and hence $f(\eaex{}_{,i})=K\eaex{}_{,i}$ with $K$ being the slope of the best fit linear function, shown as a dotted line in Fig.~\ref{fig:eta-scaling}(b). 

The presence of an excess area dependent scaling described by the slope $m$ in Fig.~\ref{fig:eta-scaling}(b) can allow one to devise strategies to estimate the range of \aex{} in cells directly from tether pulling experiments. One possible approach is to use eqn.~\eqref{eqn:mifunc} in eqn.~\eqref{eqn:etascaling} and self consistently solve for \aex{} using the relationship:
\begin{equation}
\eaex{}=\left(f(\eaex) \alpha +1 \right) {\cal G}.
\label{eqn:scfaex}
\end{equation}

Here, the variables $\alpha=\eftet{}\ertet{}/\ekbt{}$ and ${\cal G}$ are directly computed from the tether force and radius measured in  tether pulling experiments. The form of the unknown function $f(\eaex)$ is in turn obtained from simulations of model systems, that correctly accounts for the size of the cytoskeletal mesh in the target cell. The excess membrane area may then be estimated by self consistently solving eqn.~\eqref{eqn:scfaex}.

\section{Discussion} 
We have presented a computational approach based on umbrella sampling and the weighted histogram analysis technique to compute the free energy landscape and the force-extension relationship for the pulling of membrane tethers from membrane patches of different excess membrane areas, \aex{}. The tether forces measured in our simulations agree very well with \textit{in vitro} tether pulling experiments on ruptured GUVs on substrate and  on HeLa cells. Unlike existing models, we are able to account for both mechanical work as well as entropic work in tether extraction by performing finite temperature calculations, delineation of the Helmholtz free energy, and performing the analysis in an ensemble with non-zero \aex{}. Based on the computed values of the force required for tether extraction and the tether radius, we established scaling relationships involving the \ftet{}, \rtet{}, and \aex{}. We demonstrated the relevance of the calculations by showing the scaling of $\kappa$ with $\alpha$ and $\sigma$ with $\Gamma$ from the model and those obtained from 15 different cell experiments collapse on to a single curve. These scaling curves can be used to construct new schemes for estimating the excess membrane area, which  alleviate the limitations of  previous methods by being valid for large curvatures, and by taking into account the thermal membrane undulations in the high curvature limit. We have shown that our results successfully recapitulate the results of the previous model in the small-curvature limit. However, in the large-curvature limit, when the domain of applicability of the previous model is limited, we predict the values of the excess membrane areas that are substantially larger than the estimates from the small-curvature model. In light of the discussion above, there is a profound biomedical ramification of the excess membrane area distribution as revealed by our analyses of the tether pulling experiments using the fully non-linear model of the membrane patch subject to finite temperature undulations.

Our model while directly relevant to tether extraction in well behaved \textit{in vitro} setups, such as GUVs or supported bilayers, does not include the full complexity required to recapitulate the cellular experiments. The complexities arise due to: (i) the dynamic nature of the cytoskeletal reorganization, (ii) changes in \aex{} due to cellular trafficking mechanisms; the latter poses an important constraint regarding the ensemble. While in \textit{in vitro} experiments or in our model, we have the ability to either select/design a constant \aex{} or a constant $\sigma$ ensemble, it is not obvious what the correct cellular condition would be. For example, at early timescales (i.e. too short for changes in $l_{c}$) the cell membrane patch may be under a state of tension but at later times both $\sigma$ and \aex{} can change due to signaling and trafficking. Notwithstanding these considerations, our model can still be applicable under certain cellular conditions, namely (i) the timescale of the tether extraction is faster than that for cytoskeletal reorganization and trafficking  ($\sim 10$-$100$ s~\cite{Joanny:2009bs}); (ii) the dimensions of the extracted tethers are smaller than $l_c$. When these conditions are met, one can treat the tether extraction as a quasi-equilibrium process where the cytoskeleton merely serves as a pinning boundary condition for the membrane. This is further justified because the membrane tension equilibrates at a much faster time scale of $\tau_{\rm tension}=\eta_s⁄\esigt{} \sim 1$-$100$ $\mu {\rm s}$, (where $\eta_s$ is the surface dilational viscosity of the bilayer $\approx 0.35$ Ns/m~\cite{Haluska:2006fr}). Under these assumptions, \lpt{} can serve as an approximate surrogate to include cytoskeletal pinning effects. These considerations and caveats must be taken into consideration in developing experimental methods for determining \aex{} in cells based on the model we have described here.

A bi-directional coupling can be established between the cell exterior and cell interior in a ``mechano-sensitive'' fashion through the control of membrane excess area \cite{DizMunoz:2013bi}, because \aex{} is the conjugate variable for membrane tension as well as membrane curvature. Several signaling mechanic events can therefore be transduced via the regulation in \aex{} : they include cell-ECM interactions, which can tune acto-myosin tension and influence cell-proliferation through integrin-mediated signaling pathways~\cite{Mih:2012iq,Paszek:2005eh,Samuel:2011iy}. Glycocalyx remodeling can influence membrane-curvature distribution on the cell surface and initiate a proliferative cell-response, funneling through integrin-mediating signals~\cite{Paszek:2014it}. Cellular recycling pathways responsible for cargo transport from the endosome to the plasma membrane can also induce and nucleate cell-membrane protrusions providing dominant mechanisms for cell migration and motility~\cite{Zuo:2006bb,Zhao:2013hi}. These examples serve to reiterate how membrane excess area, in response to the tuning of tension, and by influencing the curvature distribution of the cell membrane, can transduce signals impacting cell-fate decisions in ECM-specific, and mechano-sensitive fashion.

 Mechanotyping cells to characterize the state of the cell membrane is, therefore, expected to be crucial in circumstances where the underlying heterogeneity is intrinsic such as in a tumor microenvironment and influences cell fate through outside-in mechanisms relayed via membrane mechanotransduction to intracellular signaling. Mechanotyping will be equally important in circumstances where the membrane plays a dominant role such as in the viral invasion of host cells in virology, formation of the immunological synapse in adaptive immunity, or targeted delivery of nanocarriers in pharmacology. 

\vspace*{10pt}

\section*{Acknowledgements}
This work was supported in part by Grants NSF-CBET-1236514 (R.R), NIH/U01EB016027 (R.R), NIH/1U54CA193417 (R.R and T.B), and NIH/R01GM097552 (T.B). T.P and S.P acknowledge support from the Wellcome Trust-DBT India alliance. Computational resources were provided in part by the Grant MCB060006 from XSEDE and NSF/DMR-1120901.
 
\section*{Author contributions statement}  
R.R. and N.R. designed and performed the simulations. A.R, T.P and S.P designed and performed the experiments. All authors were involved in data analysis and interpretation and in writing of the manuscript.
\section*{Competing financial interests} 
The authors declare that they have no competing financial interests.

\clearpage
\newpage

\renewcommand{\thefigure}{S\arabic{figure}}
\renewcommand{\thesection}{S\arabic{section}}
\renewcommand{\theequation}{S\arabic{equation}}
\setcounter{figure}{0} 
\setcounter{section}{0} 

\clearpage
\newpage
\vspace*{\fill}
\begingroup
\centering
\begin{center}
	\Large{Supplementary Information}
	\end{center}
\endgroup
\vspace*{\fill}

\clearpage
\newpage

\section{Dynamical Triangulated Monte Carlo}
The dynamical triangulation Monte Carlo technique consists of two independent moves to alter the degrees of freedom that define the triangulated surface which is taken as a model for the fluid membrane~\cite{Kroll:1992vh,Ramakrishnan:2010hk}: \\

\begin{figure}[!h]
	\centering
	\includegraphics[width=12cm,clip]{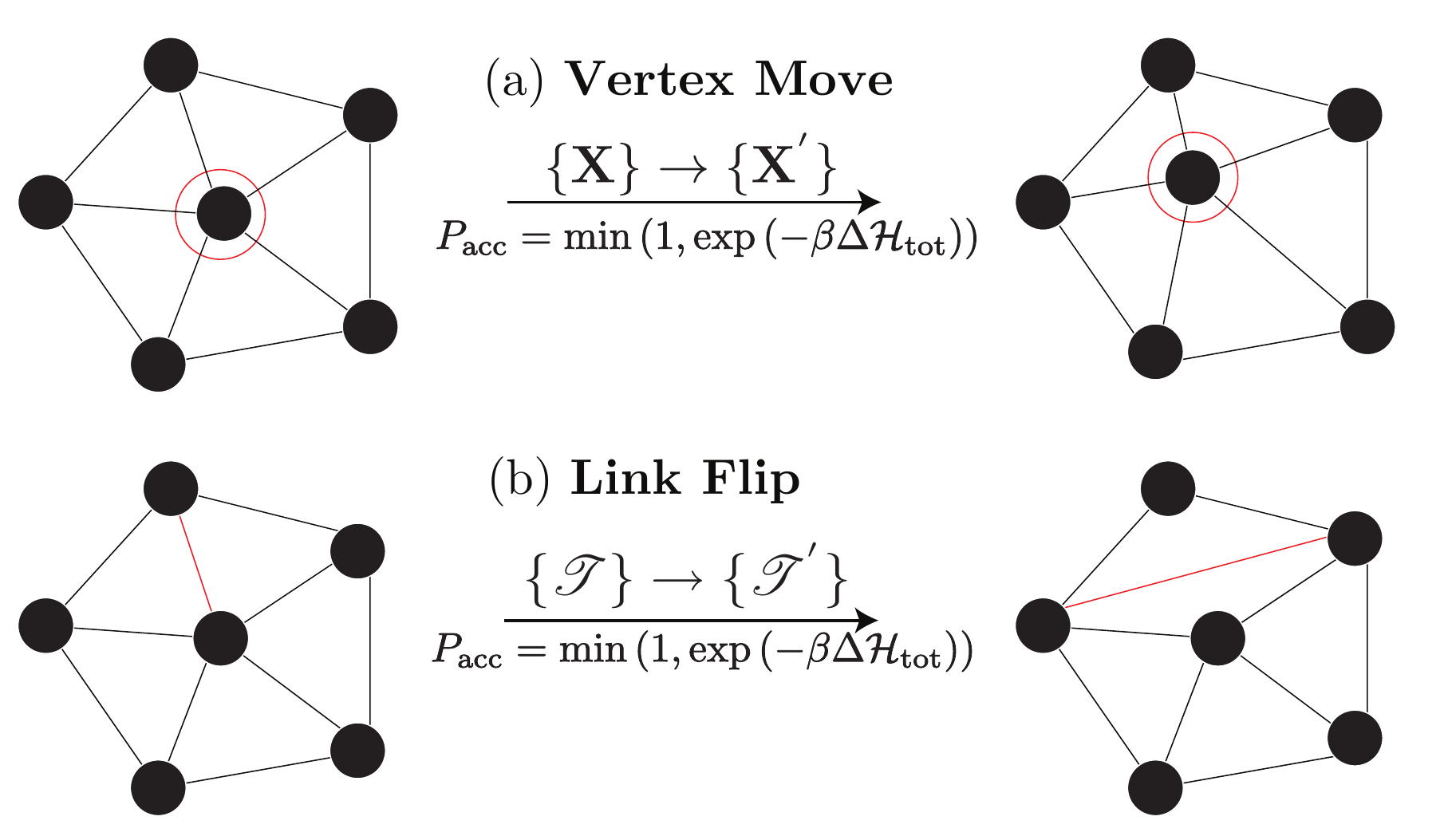}
	\caption{\label{fig:mcs} Dynamical triangulated Monte Carlo scheme to independently modify the position (a) and the connectivity (b) of the vertices in the triangulated surface model.}
\end{figure}

	{\bf 1) Vertex Move:} A randomly chosen vertex is randomly displaced to a new position within a cube of size $\epsilon$, centered around the  vertex. The move is performed by the  holding the connectivity fixed as shown in Fig.~\ref{fig:mcs}(a) and accepted using the Metropolis scheme~\cite{Metropolis:1953in}.  \\
	
	{\bf 2) Link Flip:} A randomly chosen tether shared between two triangles on the surface is removed and reconnected between the two previously unconnected vertices as shown in Fig.~\ref{fig:mcs}(b), by holding the vertex positions fixed. \\
	
Both moves are accepted using the standard Metropolis scheme with a probability given by the Boltzmann constant of the energy change ($\Delta {\cal H}_{\rm tot})$ due to the move. In the case of tether pulling simulations the total energy of the membrane is given by ${\cal H}_{\rm tot}={\cal H}+{\cal H}_{\rm bias}$, where ${\cal H}$ denotes the elastic Hamiltonian and ${\cal H}_{\rm bias}$ is the harmonic biasing potential as defined in the main manuscript. Here, \kbt{} = 1 is the inverse temperature, with $k_{\rm B}$ the Boltzmann constant and $T$ the absolute temperature. \\

The state of the membrane can be affected by variations either in the bending stiffness  or in the self-avoidance parameter, leading to membranes with different excess areas \aex{}.  Snapshots of the membrane conformations in the parameter space of bending rigidity and excess area are shown in Fig.~\ref{fig:conformations}.
\clearpage
\newpage

\section{Membrane conformations in various limits}
\begin{figure}[!h]
	\centering
	\includegraphics[width=12cm,clip]{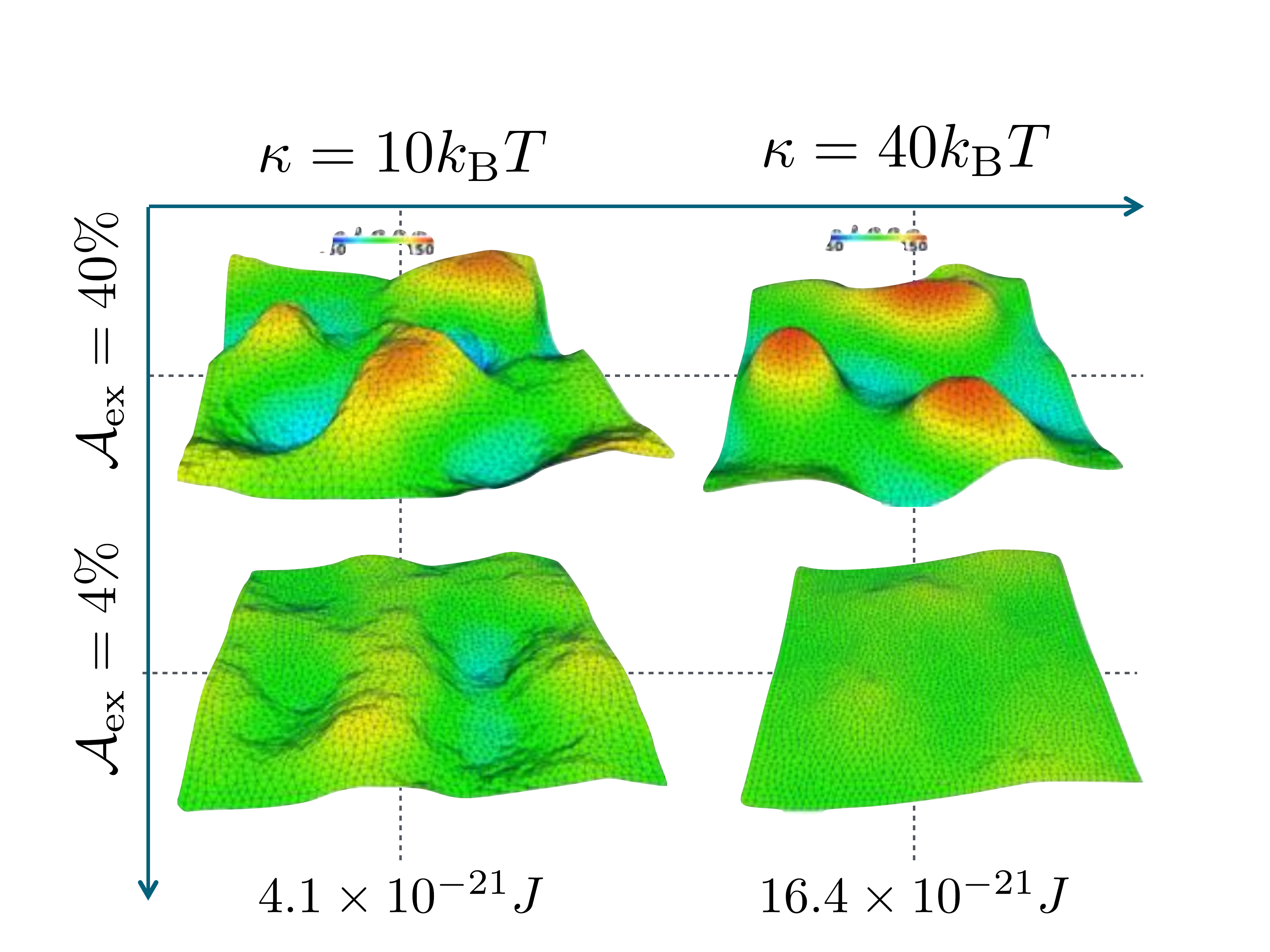}
	\caption{\label{fig:conformations} Conformations of membranes with different bending stiffness and excess area. Shown are shapes for two values of the excess area $\eaex{}=4$ and  $40\%$.
	}
\end{figure}

The conformations of a planar membrane, when ${\cal H}_{\rm bias}=0$, for two different bending rigidities ($\kappa=10$ and $40$ \kbt{}) for two different values of \aex{} ($=4\%$ and $40\%$) are shown in Fig.~\ref{fig:conformations}. The surface is colored with respect to the $z$ position of the vertices.

\clearpage
\newpage

\section{Undulation spectrum for the planar membrane}
In the continuum limit, a planar membrane can be parameterized based on its height with respect to a reference plane and such a parameterization is called the Monge gauge. If the reference plane is taken to be the  plane, then the height of the membrane at a chosen point on the plane, with coordinates $x$ and $y$, is given by $h(x,y)$.  The height of the membrane can also be expressed in terms of its Fourier modes as~\cite{Seifert:1997wq}
\begin{equation}h({\bf X})= \frac{1}{{\cal L}_{\rm patch}^2} \int d{\bf q} \,\, h_{\bf q} \exp(-i{\bf q} \cdot {\bf X})
\label{eqn:fourier-h}
\end{equation}

\begin{figure}[!h]
	\centering
	\includegraphics[width=12cm,clip]{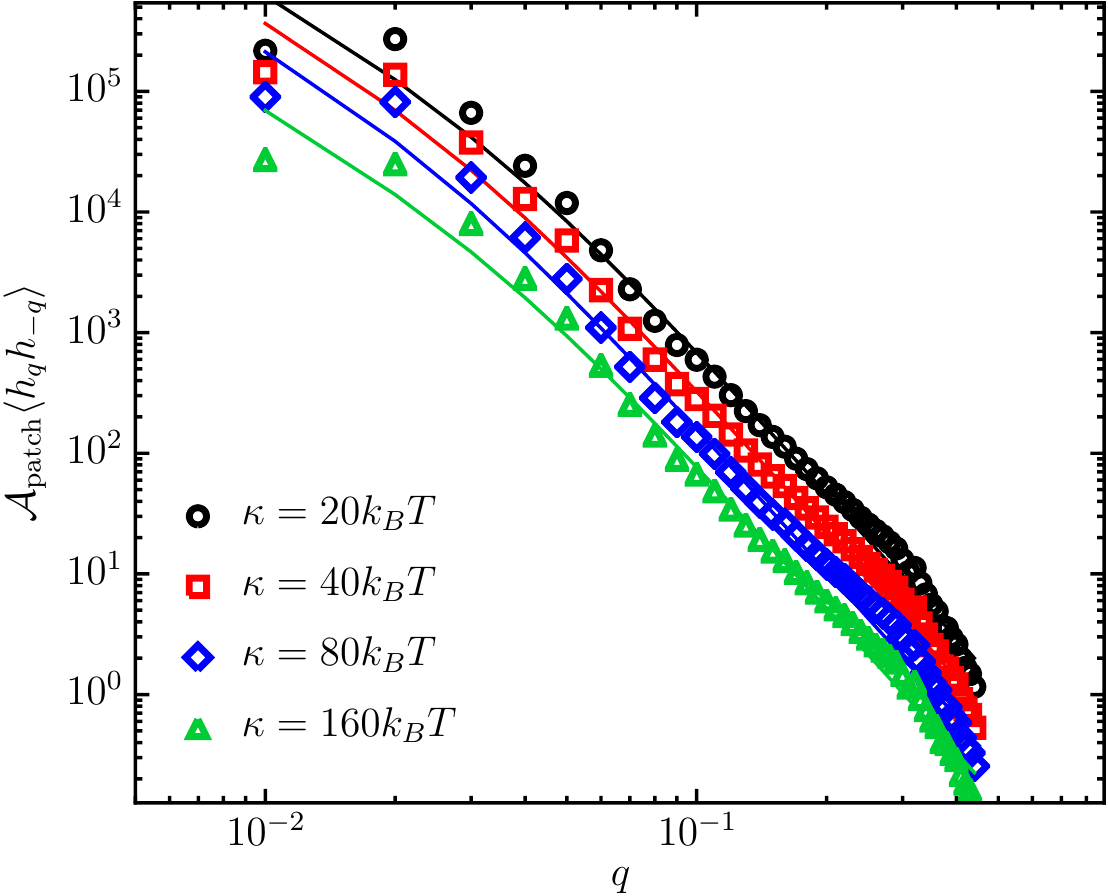}
	\caption{\label{fig:fluctspect} Validation of the small deformation limit. The power spectrum, for each of the Fourier modes, scales as $q^{-4}$  when the membranes have small excess area or large bending stiffness.}
\end{figure}

Here we have used the short hand notations ${\bf X}=[x,y]$ and ${\bf q}=[q_{x},q_{y}]$  to denote two dimensional real and Fourier spaces and the Fourier amplitude also has two components given by $h_{\bf q}=[h_{q_{x}},h_{q_{y}}]$.  When the elastic Hamiltonian ${\cal H}$ (see eqn. 1 of the main manuscript) is expressed in terms of its Fourier modes, the power spectrum for each of the modes can be shown to obey the relation,
\begin{equation}
\eapt{} \left \langle h_q h_{-q} \right \rangle = \dfrac{\ekbt{}}{\kappa q^4+\sigma q^2}
\label{eqn:hqhq}
\end{equation}
This result is derived for nearly planar membranes (where $|\nabla h \ll 1|$) and hence should be reproducible in the simulations for membranes with either large bending stiffnesses or small excess areas or both. The power spectrum for planar membranes with small excess area and for a range of values of  is shown in Fig.~\ref{fig:fluctspect}. The observed undulation modes scale as $q^{-4}$, which is in good agreement with the theoretical expression given above. However, it should be remembered that membranes with large excess area would not adhere to this scaling behavior, since the excess area manifests as large amplitude undulations, which takes the systems beyond the small deformation limit (as $|\nabla h \sim 1|$).

\clearpage
\newpage
\section{Properties of the tether as a function of $\kappa$ and \aex{}}
\begin{figure}[!h]
	\centering
	\includegraphics[width=15cm,clip]{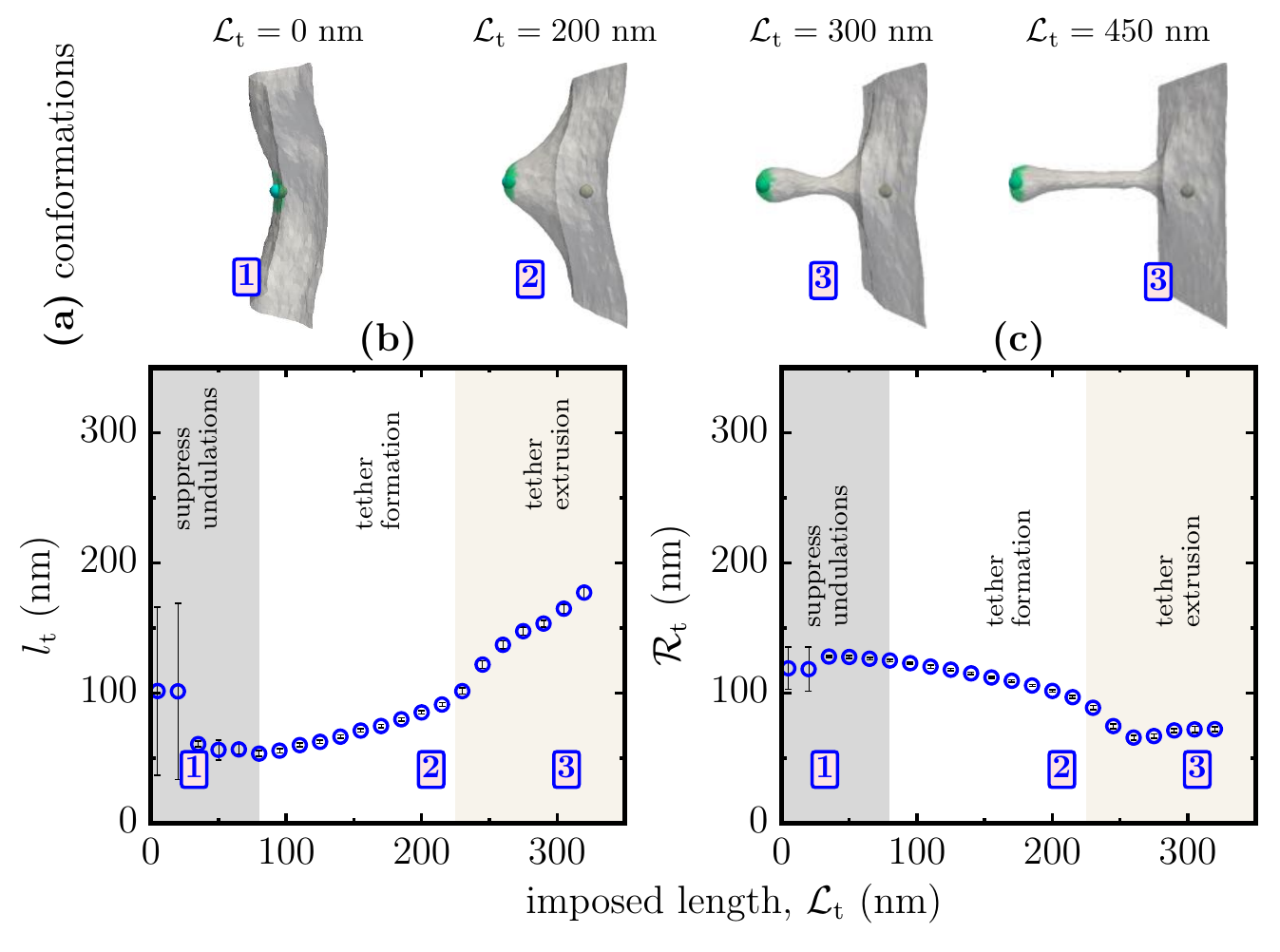}
	\caption{\label{fig:lowaex} The length and radius of the tether extracted from a membrane with \kap{20} and \exarea{10} as a function of the imposed tether length \Ltet{}.}
\end{figure}

In this section, we display the effect of the membrane excess area and bending rigidity on the length and radius of a tether extracted from a cell membrane. In Fig.~\ref{fig:lowaex} we show \ltet{} and \rtet{}, along with the membrane conformations, as a function of the imposed tether length \Ltet{} for a membrane with \kap{20} and \exarea{10}. 

\begin{figure}[!h]
	\centering
	\includegraphics[width=15cm,clip]{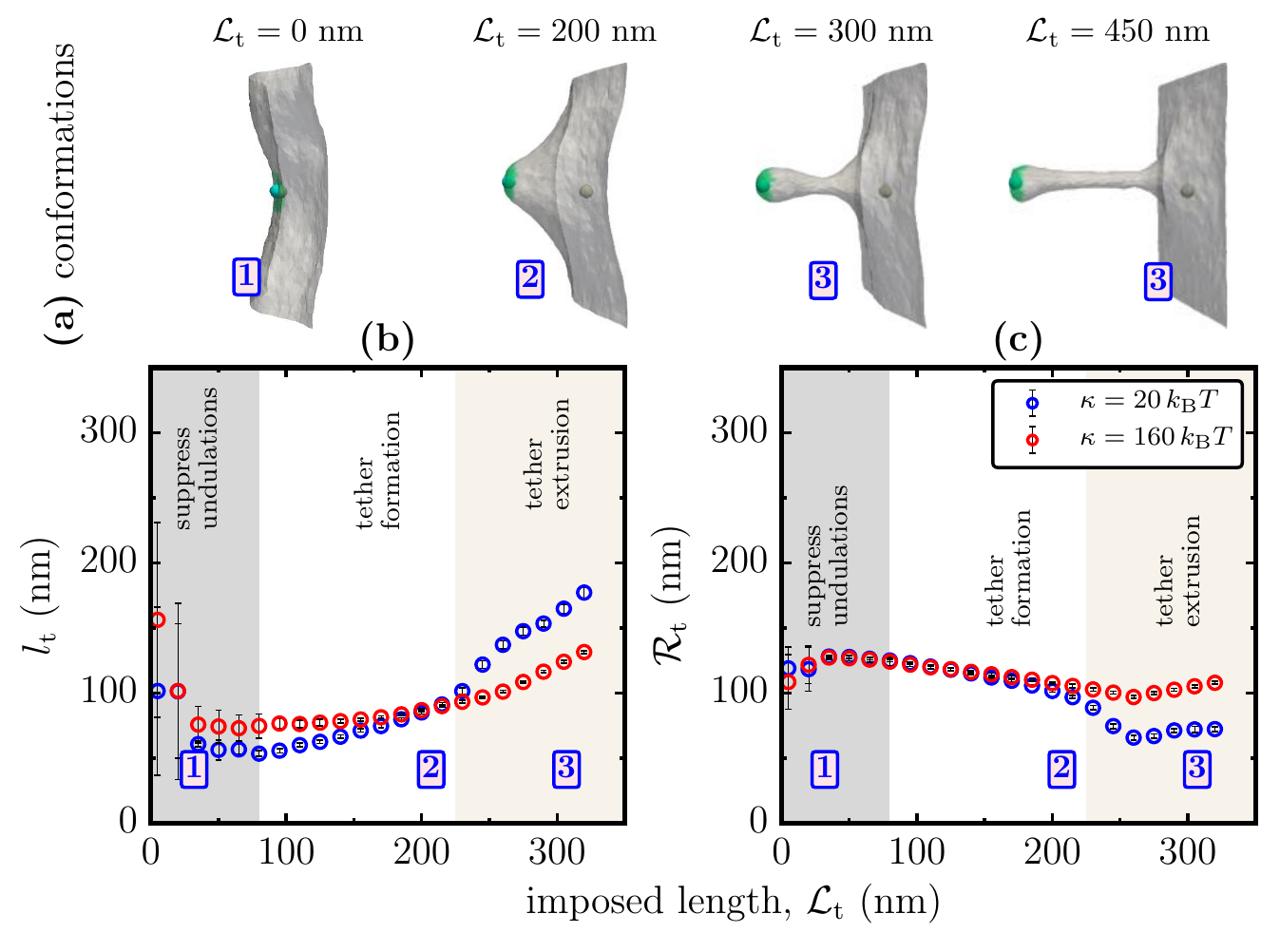}
	\caption{\label{fig:effK} Effect of $\kappa$ on the length and radius of the extracted tether as a function of the imposed tether length \Ltet{}, for membranes with similar excess areas, taken to be \exarea{10}.}
\end{figure}

Similarly, in Fig.~\ref{fig:effK} we show the effect of $\kappa$ on \ltet{}
and \rtet{} for membranes with similar excess areas, chosen to be \exarea{10}. The tether pulling data is displayed for $\kappa=20$, and $160$ \kbt{}.

As noted in the discussions on Fig.2 in the main manuscript, we find both the systems to exhibit the three distinct scaling regimes previously identified for the tether radius. However, for the membranes with low excess area considered here we find the third regime to occur at a smaller value of \Ltet{} compared to that seen for membranes with large excess areas. Similarly, the value of \rtet{} in the final regime is an increasing function of $\kappa$, as is evident from Fig.~\ref{fig:effK}.

\clearpage
\newpage

\section{Tether pulling experiments}
\begin{figure}[!h]
	\centering
	\includegraphics[width=15cm,clip]{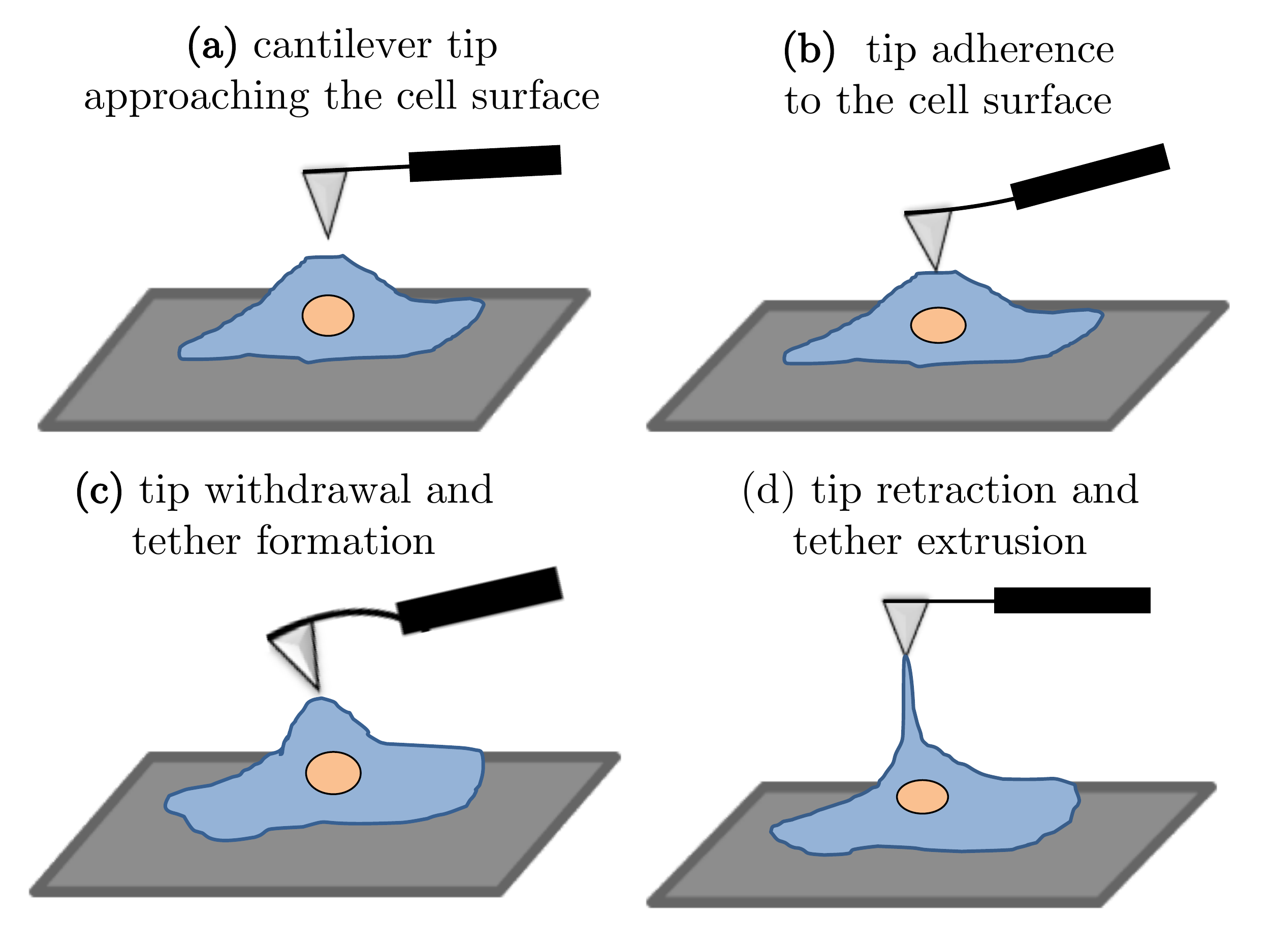}
	\caption{\label{fig:exptether} Various stages of a tether pulling experiment.}
\end{figure}

A typical tether pulling experiment proceeds through many stages as illustrated in Fig.~\ref{fig:exptether}. In the first stage, the tip of an atomic force microscope (AFM), attached to a cantilever, is indented into the cell surface and held fixed until the tip makes a contact with the cell membrane; these stages are illustrated in Figs.~\ref{fig:exptether}(a) and (b). Stage (b) in the experiments is analogous to the initial configurations used in our simulations. After the formation of a stable contact the AFM tip is retracted at a constant velocity until it returns to its undeflected state, as shown in Figs.~\ref{fig:exptether}(c) and (d). In the course of retraction the adherence between the tip and the membrane leads to formation of a tether followed by its extrusion and these process are identical to those observed in our simulations and described in Sec.4 of the main manuscript.

\clearpage
\newpage
\section{Mechanical properties of the 15 different cells in the CNS}
\begin{figure}[!h]
	\centering
	\includegraphics[width=15cm,clip]{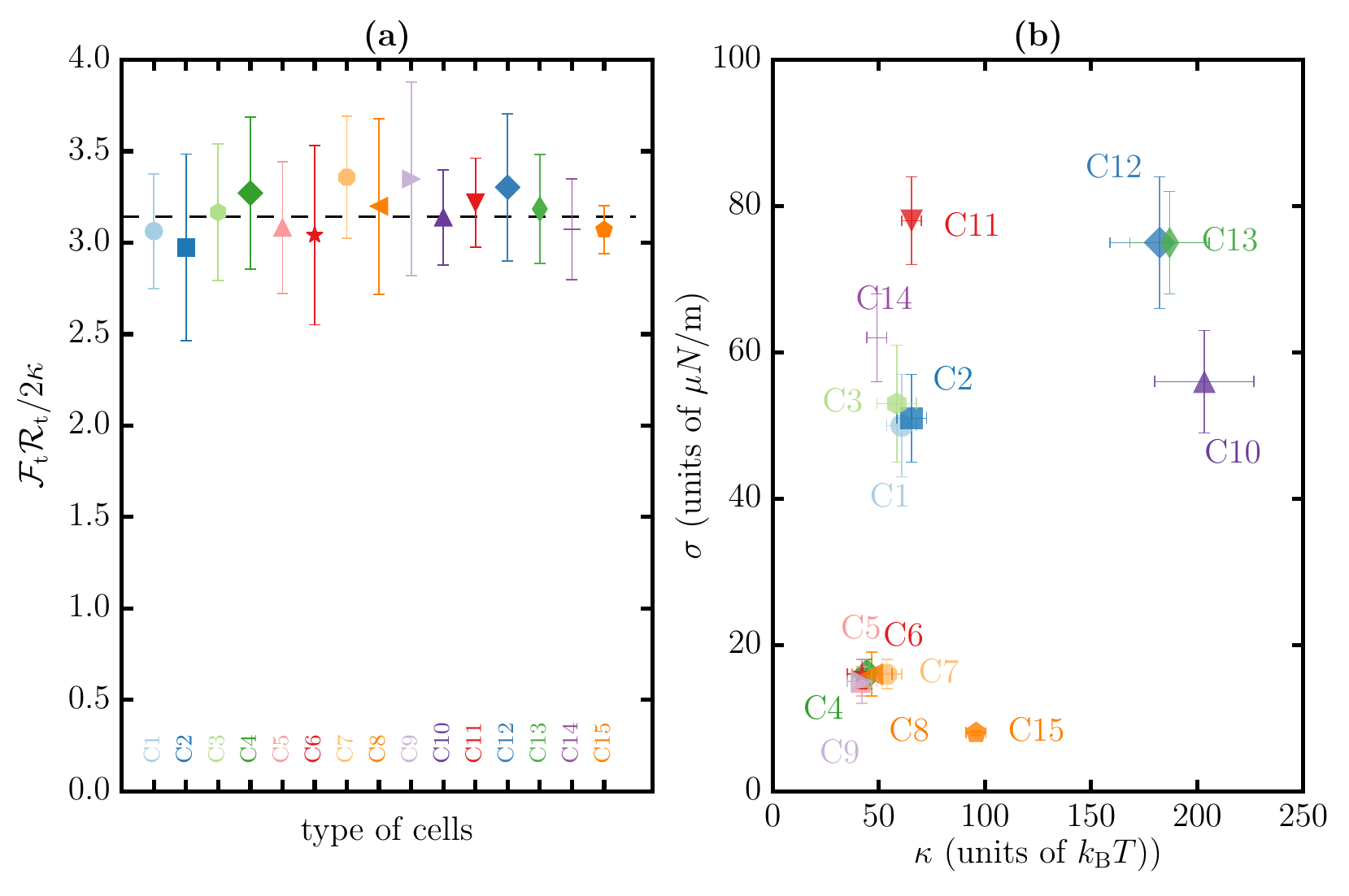}
	\caption{\label{fig:15cns} (a) The scaling relation $\eftet\ertet/2\kappa$ and (b) the values of $\kappa$ and $\sigma$ for 15 different cells (marked C1--C15) in the CNS. Data from Pontes et. al.~\cite{Pontes:2013gl}.}
\end{figure}

Here we show data from Pontes et. al.~\cite{Pontes:2013gl} for the mechanical properties of 15 different cells in the central nervous system (CNS). The tether force \ftet{} and radius \rtet{} for each of these cells (marked C1--C15) satisfies the scaling relation $\eftet\ertet/(2\kappa)=\pi$ and this is shown in Fig.~\ref{fig:15cns}(a). The values of $\kappa$ and $\sigma$ are shown in  Fig.~\ref{fig:15cns}(b) and the spread of the data show three characteristic mechanical regimes namely: (i)low $\kappa$ and low $\sigma$, (ii)low $\kappa$ and high $\sigma$, and (iii) high $\kappa$ and high $\sigma$.

\clearpage
\newpage

\section{Movie M1}
The movie shows the conformations of a tether extracted from a planar membrane as a function of the reaction coordinate \Ltet{} -- data shown for a membrane with $\elpt{}=510$ nm, $\kappa=40$ \kbt{}, and \exarea{40}. The histogram shown alongside corresponds to  the distribution of the mean curvature of the membrane surface
\begin{figure}[!h]
	\centering
	\includegraphics[width=12.5cm,clip]{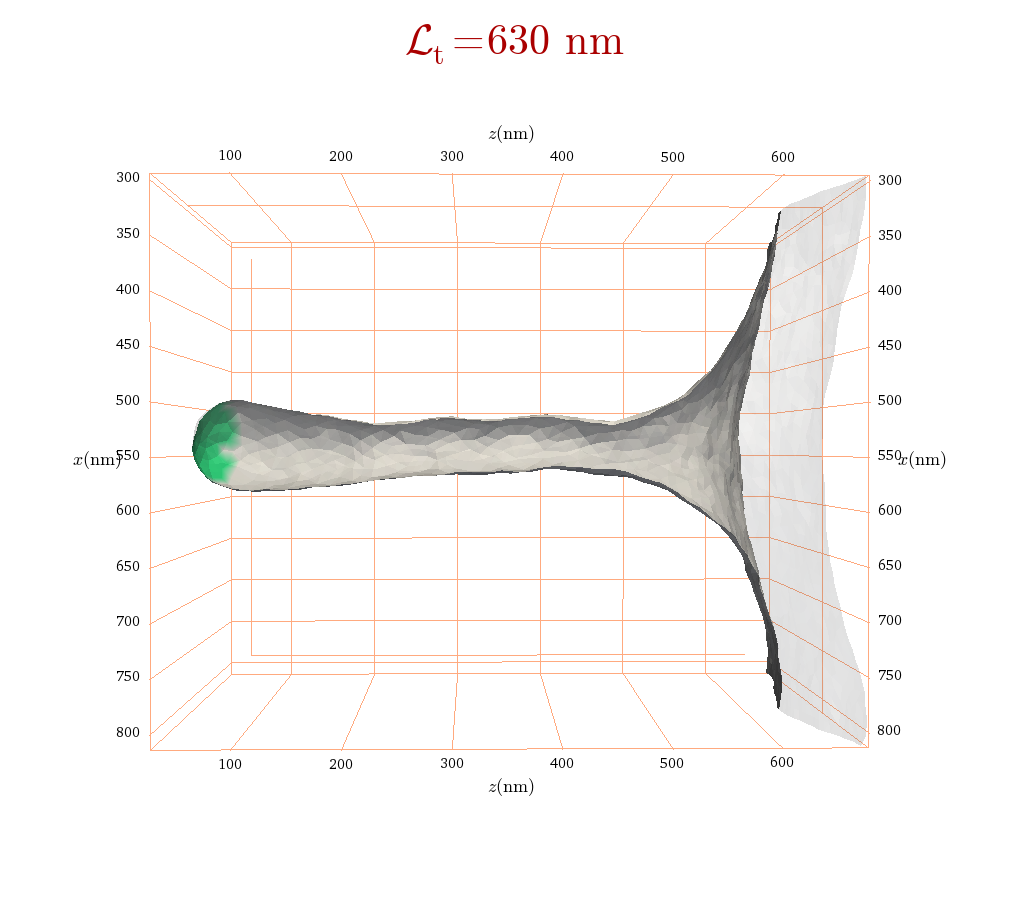}
	\caption{Movie showing the evolution of tether as a function of the reaction coordinate \Ltet{}.}
\end{figure}

\end{document}